\documentclass[structabstract]{aa}  
\usepackage{graphicx}
\usepackage{txfonts}
\usepackage{longtable,lscape}
\usepackage[round]{natbib}

\begin{document}

   \title{Revisiting the shocks in BHR71:~new observational constraints and H$_2$O predictions for \textit{Herschel}}

   \author{A. Gusdorf,
          \inst{1}
          T. Giannini,
          \inst{2}
          D. R. Flower,
          \inst{3}
          B. Parise, 
          \inst{1}      
          R. G\"usten,
          \inst{1}
          \and
          L. E. Kristensen
          \inst{4}
          }

   \institute{Max Planck Institut f\"ur Radioastronomie, 
              Auf dem H\"ugel, 69, 53121 Bonn, Deutschland\\
              \email{agusdorf@mpifr-bonn.mpg.de}
         \and
              INAF-Osservatorio Astronomico di Roma, Via di Frascati 33, I-00040 Monteporzio Catone,
Italy 
	\and
	    Physics Department, The University, Durham DH1 3LE, UK
         \and
             Leiden Observatory, Leiden University, Niels Bohrweg 2, 2333 CA Leiden, The Netherlands  \\
             }
   \date{}

 
  \abstract
   {During the formation of a star, material is ejected along powerful jets that impact the ambient material. This outflow phenomenon plays an important role in the regulation of star formation. Understanding the associated shocks and their energetic effects is therefore essential to the study of star formation.}
   {We present comparisons of shock models with observations of H$_2$ and SiO emission in the bipolar outflow BHR71, and predict water emission, under the basic assumption that the emission regions of the considered species coincide, at the resolution of currently available observations.} 
   {New SiO observations from APEX are presented, and combined with \textit{Spitzer} and ground-based observations of H$_2$ to constrain shock models. The shock regions associated with targeted positions in the molecular outflow are studied by means of a 1D code that generates models of the propagation of stationary shock waves, and approximations to non-stationary ones.} 
   {The SiO emission in the inner part of the outflow is concentrated near the apex of the corresponding bow-shock that is also seen in the pure rotational transitions of H$_2$. Simultaneous modelling is possible for H$_2$ and SiO and leads to constraints on the silicon pre-shock distribution on the grain mantles and/or cores. The best-fitting models are found to be of the non-stationary type, but the degeneracy of the solutions is still large. Pre-shock densities of 10$^4$ and 10$^5$~cm$^{-3}$ are investigated, and the associated best-model candidates have rather low velocity (respectively, 20-30 and 10-15 km s$^{-1}$) and are not older than 1000~years. We provide emission predictions for water, focusing on the brightest transitions, to be observed with the PACS and HIFI instruments of the \textit{Herschel} Telescope.}
   {}

  \authorrunning{A. Gusdorf et al.}
   \maketitle
%

\section{Introduction}

The propagation of shock waves is an ubiquitous phenomenon in the interstellar medium, where the sound speed is low due to the temperature conditions. Observations over the past few decades (\citealt{Snell80}) have shown that, at the early stages of the formation of solar-type stars,  the process of mass accretion is almost always associated with mass ejection, in the form of collimated jets, extending from a few astronomical units up to parsecs from the exciting source (\citealt{Dougados00,Bally96}). The supersonic impact between the jet and the parent cloud generates a shock front, which propagates in the collapsing interstellar gas, and also an inverse shock that propagates along the jet itself. Large cavities, called bipolar outflows, appear, which have been extensively studied through the molecular emission that they generate (\citealt{Bachiller96}). At the apices of these cavities, the shock wave heats, compresses and accelerates the ambient interstellar gas. As the temperature rises to at least a few thousands degrees, the energy barriers of numerous reactions involving neutral or ionized molecules can be overcome and the chemistry of certain species can be significantly altered (\citealt{Bachiller01}). Similarly, processes specific to the propagation of shock waves affect the interstellar grains (\citealt{Flower031,Guillet07,Guillet09}), leading to the injection of molecular and atomic species into the gas phase, such as silicon-bearing species (\citealt{Schilke97,Gusdorf081}, hereafter G08a). The time-scales involved in the heating and in some of the shock chemical processes are short (10$^2$ to 10$^4$ years; \citealt{Gusdorf082}, hereafter G08b), so the shocked region rapidly acquires a chemical composition distinct from that of the quiescent medium. As the gas cools down radiatively, reactions with high energy barriers no longer operate, and the chemistry is dominated again by low-temperature reactions. The resulting molecular emission can be used as a diagnostic tool to study the physical and chemical characteristics of the shocked region. 

Recently, G08b placed constraints on shock parameters using H$_2$, SiO and CO observations in the B1 knot of the L1157 outflow. At the same time, the HIFI and PACS receivers on \textit{Herschel} have observed the entire L1157 outflow (\citealt{Nisini10}) and provided velocity-resolved data of the L1157-B1 knot (\citealt{Lefloch10,Codella10}). Their studies indicate that the H$_2$, SiO and H$_2$O emission regions are spatially coincident on large scales. They also shed light on the necessity to provide reliable models of water emission in order to better understand its production in shocks and its role in the cooling of the outflow. Recently, \citet{Flower10} (hereafter FP10) presented an extensive study of water emission for a small grid of stationary shock models, and compared it to \textit{Spitzer} observations of water emission in NGC2071, with a certain level of agreement. Our aim is to push the confrontation between models and observations one step further, by using the models, which have already demonstrated their ability to reproduce some observations, to predict water line intensities. Using an approach to comparing models with observations that is similar to that of G08b, as applied to H$_2$ and SiO, we aim to constrain the physical and chemical parameters of the targeted region. In a second step, we incorporate these preliminary constraints and the latest compilation of molecular data for H$_2$O in an LVG radiative transfer code to provide predictions of water emission, under the assumption that the emission regions of the various species coincide.

The choice of target is of crucial importance, as our candidate must have been the object of prior studies, and must be a good potential target for contemporary and future instruments such as \textit{Herschel} and ALMA. Located at a distance of about 175 pc, the Bok globule BHR71 (\citealt{Bourke95}) is a well-known example of isolated star formation observable in the southern hemisphere, and hence is a reasonable candidate for future ALMA observations. Its observation by the \textit{Herschel} telescope is guaranteed through the key-program WISH, and also DIGIT. The associated outflow region contains signatures of two distinct molecular jets (\citealt{Bourke97,Myers98,Parise06}) centered on two different protostellar sources, IRS1 and IRS2 (\citealt{Bourke012}) separated by $\sim$3400~AU. Bright HH objects are associated with the blue-shifted lobes of each outflow, HH320 and HH321 (\citealt{Corporon97}). They have been imaged in the \ion{S}{ii} transition at 6711~\AA, indicating that at least part of the outflows are dissociative. The dynamical age of HH321 is estimated to be 400~years (\citealt{Corporon97}). It has not yet been possible to determine the dynamical age for HH320. NIR JHK-band spectra covering both HH320 and HH321 were obtained by \citet{Giannini04}. Through detailed shock modelling they derived a pre-shock density of 10$^4$~cm$^{-3}$ and shock velocity of 41~km/s for HH320A; this was found by fitting a non-steady-state shock model to the observed H$_2$ lines. The age of the non-steady-state shock is 475~years, which is more or less in agreement with the dynamical age of the HH321 flow. The  pre-shock density is lower than is predicted on the basis of CO observations, 10$^5$~cm$^{-3}$ (\citealt{Parise06}). However, the latter is the density of the molecular outflow, which is compressed relative to the ambient pre-shock cloud. These inner parts of the bipolar outflow, covering the HH objects, have also been targeted by the  \textit{Spitzer} Space Telescope, and the corresponding results in terms of rotational H$_2$ excitation diagrams have been made available by \citet{Neufeld09}. Another article is dedicated to the parameters map analysis based on these data (\citealt{Giannini11}). Finally, \citet{Garay98} have presented maps of SiO~(2--1) and (3--2) of the whole bipolar outflow.

In Section~\ref{sec:h2oam}, we presented a combined analysis of both rovibrational transitions of H$_2$ from \citet{Giannini04} and rotational lines, as compiled by \textit{Spitzer} (\citealt{Neufeld09}). We then briefly present the code that is used to simulate the propagation of shock waves through the interstellar medium, and introduce the grid of models that we computed. Finally we describe the constraints that we obtain in several knots of the BHR71 outflow by comparing more or less complete H$_2$ observations with models. In Section~\ref{sec:SiOoam}, we present the results of comparisons between observations of SiO and the models. We describe the observing setups, the actual observations, and the results in terms of simultaneous fits of the H$_2$ and SiO molecules. Section~\ref{sec:wep} contains our predictions of water emission line intensities. We briefly summarize our methodology and then discuss the appropriateness of the steady-state and statistical equilibrium assumptions, before presenting a comparison of results obtained using each of these approximations. Section~\ref{sec:cr} contains our concluding remarks.


\section{H$_2$ observations and modelling}
\label{sec:h2oam}

\subsection{Observations}
\label{sub:h2o}

There are two distinct sets of H$_2$ observations. The details of the ground-based observations of the rovibrational transitions are reported in \citet{Giannini04}. The \textit{Spitzer} observations of the pure rotational transitions, described in \citet{Neufeld09}, are the subject of an separate article (\citealt{Giannini11}). 

The comparison between observations of H$_2$ emission and model results is usually based on excitation diagrams that are derived for selected emission regions. Such excitation diagrams display ln($N_{vj}/g_j$) as a function of $E_{vj}/k_{\rm B}$, where $N_{vj}$ (cm$^{-2}$) is the column density of the rovibrational level ($v, J$), $E_{vj}/k_{\rm B}$ is its excitation energy (in K), and $g_j = (2j+1)(2I+1)$ its statistical weight (with $I=1$ and $I=0$ in the respective cases of ortho- and para-H$_2$). If the gas is thermalized at a single temperature, all the points in the diagram fall on a straight~line.

\begin{table}
\caption{The considered positions in the BHR71 outflow.}             
\label{table1}      
\centering                          
\begin{tabular}{c c c}        
\hline            
position name & R. A. (J2000)  & Decl. (J2000)  \\    
\hline \hline         
   central position & 12$^{\rm h}$01$^{\rm m}$36 \fs 07 & $-$65$^\circ$08$'$50 \farcs 5 \\   
   IRS1 & 12$^{\rm h}$01$^{\rm m}$36 \fs 63 & $-$65$^\circ$08$'$48 \farcs 5 \\   
   IRS2 & 12$^{\rm h}$01$^{\rm m}$34 \fs 00 & $-$65$^\circ$08$'$44 \farcs 5 \\            
   HH320AB & 12$^{\rm h}$01$^{\rm m}$31 \fs 20 & $-$65$^\circ$08$'$05 \farcs 0 \\      
   HH321AB & 12$^{\rm h}$01$^{\rm m}$36 \fs 70 & $-$65$^\circ$09$'$32 \farcs 0 \\
   knot 5 & 12$^{\rm h}$01$^{\rm m}$32 \fs 60 & $-$65$^\circ$08$'$23 \farcs 3 \\
   knot 6 & 12$^{\rm h}$01$^{\rm m}$36 \fs 00 & $-$65$^\circ$09$'$18 \farcs 6 \\
   knot 8 & 12$^{\rm h}$01$^{\rm m}$38 \fs 90 & $-$65$^\circ$10$'$07 \farcs 0 \\
   SiO peak & 12$^{\rm h}$01$^{\rm m}$33 \fs 80 & $-$65$^\circ$07$'$24 \farcs 0 \\
   SiO knot & 12$^{\rm h}$01$^{\rm m}$36 \fs 00 & $-$65$^\circ$07$'$34 \farcs 0 \\
\hline                                  
\end{tabular}
\end{table}

In their original study of conditions in the BHR71 outflow, \citet{Giannini04} extracted the excitation diagrams corresponding to nine positions (knots 1 to 9) and four HH objects (HH320A\slash B and HH321A\slash B), in order to compare their observations with shock models. In the present study, we attempt to make use of most of the available data, both in the pure rotational and in the rovibrational transitions of molecular hydrogen. Unfortunately, combining the two sets of observations is only possible where both sets of data are available and when a cross calibration can be performed to ensure their consistency. We consequently make use of the following excitation diagrams~: 
\begin{itemize}
\item pure rotational transitions for the \lq H$_2$ knots' 5, 6, 8 as denoted by \citet{Giannini04}~;
\item rotational and rovibrational transitions for the two Herbig-Haro objects that are called HH320AB and HH321AB, as defined by \citet{Giannini11}~;
\item pure rotational transitions on the \lq SiO knot' (see Subsection~\ref{sub:SiOr}).
\end{itemize}

   \begin{figure}
   \centering
   \includegraphics[height=9cm,angle=-90]{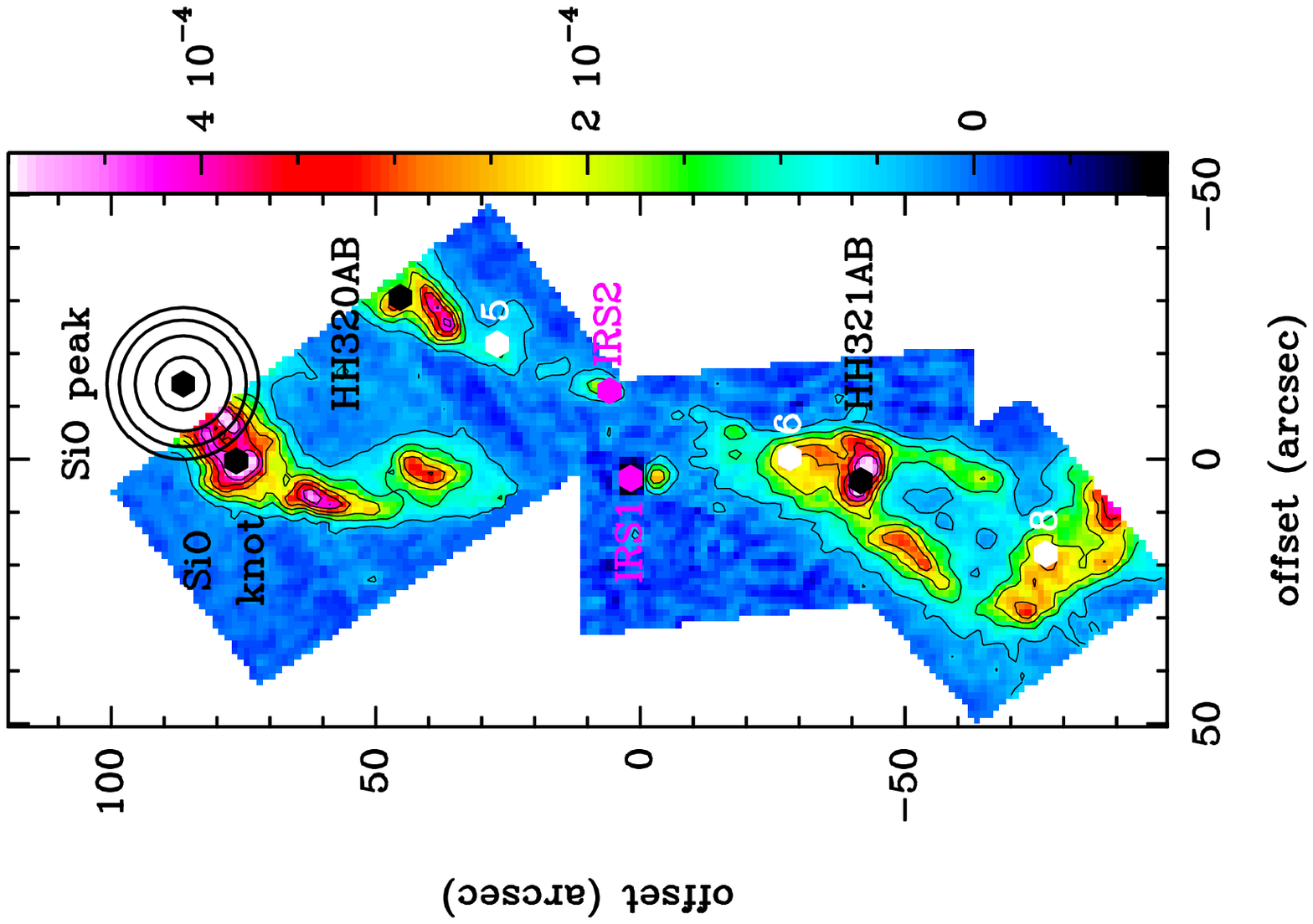}
      \caption{The internal parts of the BHR71 outflow as observed by \textit{Spitzer} in the H$_2$ 0--0 S(5) transition (\citealt{Neufeld09}); the colour scale unit is erg~cm$^{-2}$~s$^{-1}$~sr$^{-1}$. The contours correspond to the levels of 3, 9, 15, 21, 27, and 33~$\sigma$. The pink hexagons give the positions of the IRS1 and IRS2 sources as determined by \citet{Bourke01}. The positions of the Herbig-Haro objects HH320AB and HH321AB are plotted in black as well as those of the SiO peak (referring to the point of maximum SiO~(5--4) emission, see Section~\ref{sec:SiOoam}), and of the SiO knot, which was chosen to perform our simultaneous fit of H$_2$ and SiO emission, and also to make predictions of water emission. The other emission knots that were analysed are plotted in white (knots 5, 6, and 8, following the labelling of  \citealt{Giannini04}). Around the SiO-peak, four radii are displayed, corresponding to 5, 9, 12, and 14.4$''$. The first represents the size of our region of analysis. The latter three correspond to the beam of the APEX telescope for the SiO (8-7), (6-5) and (5-4) transitions, respectively; see text.}
         \label{figure1}
   \end{figure}

The corresponding list of coordinates is provided in Table~\ref{table1}. All these points are presented in Figure~\ref{figure1}, which shows the internal parts of the BHR71 outflow, as mapped by the \textit{Spitzer} Space Telescope in the 0--0~S(5) spectral line of H$_2$ (\citealt{Neufeld09}). In each case, the excitation diagrams were extracted for a circular emission region of 5$''$ radius, centered on these positions, shown for the case of the so-called \lq SiO peak' (see Subsection~\ref{sub:SiOr}) in Figure~\ref{figure1} (smallest circle). In the case of the Herbig-Haro objects, the consistency between the rotational and rovibrational data sets was ensured by means of the following method, as explained by \citet{Giannini11}. First, the image of H$_2$ 1--0 S(1) line at 2.122~$\mu$m (retrieved from the ESO archive at http://archive.eso.org/eso/eso\_archive\_main.html) is degraded to the \textit{Spitzer} spatial resolution of 2$^{\prime\prime}$/px. Then, this image is calibrated by using 2MASS fluxes of several bright stars in the field, and the photometry is computed in the same areas of 5$^{\prime\prime}$ radius as the \textit{Spitzer} images. To find the intercalibration factor between imaging and spectroscopy, the measured 2.12~$\mu$m flux was compared with the flux measured in the 1$^{\prime\prime}$ slit where spectroscopic observations have been performed. Finally, the same factor was applied to the fluxes of the other NIR lines, adopting averages of the line ratios (relative to the 2.12~$\mu$m transition) in the A and B substructures of HH320 and HH321.
   
\begin{figure*}
   \centering
   \includegraphics[width=18cm]{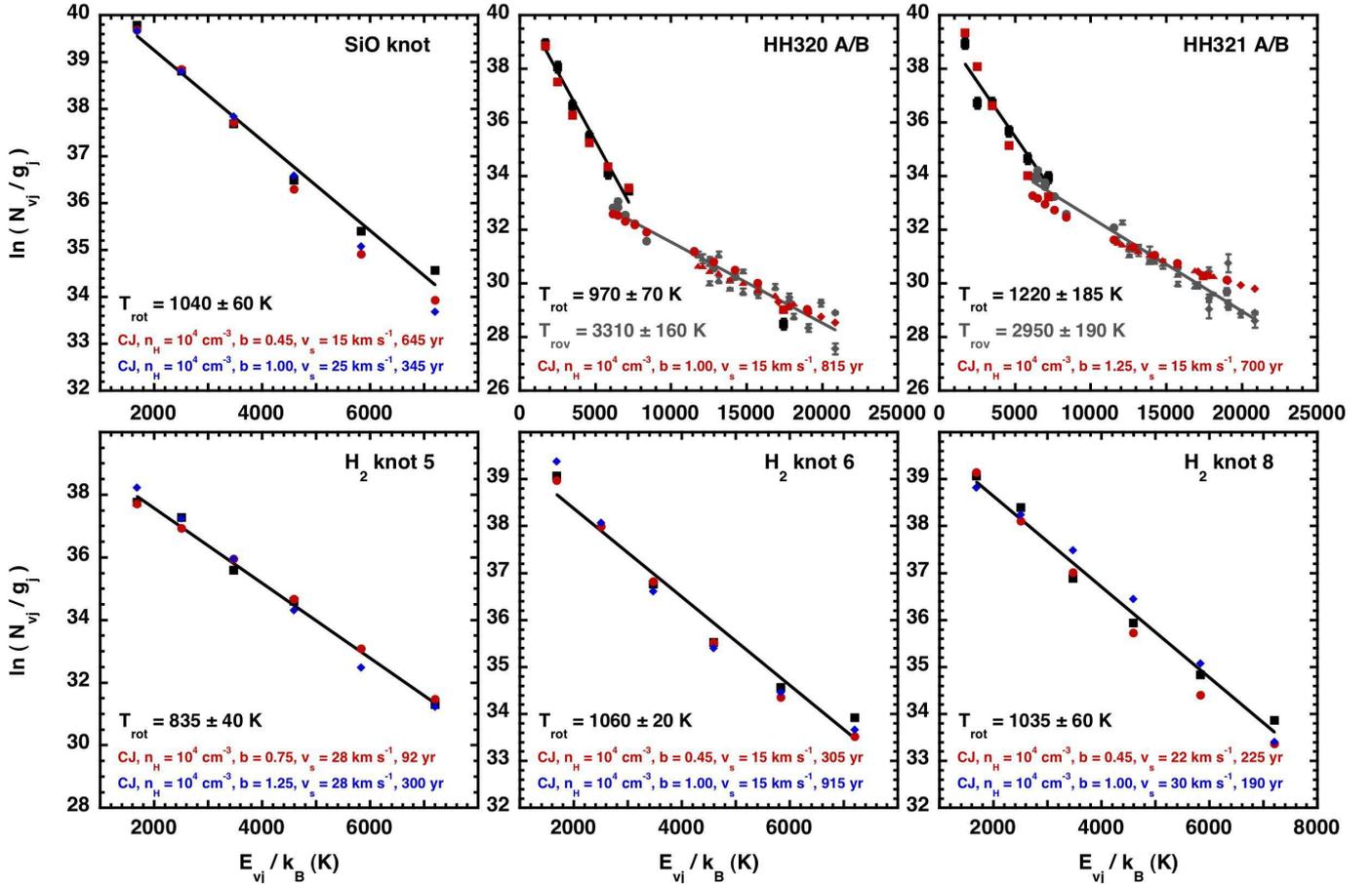}
      \caption{The H$_2$ excitation diagrams observed for each position indicated in Figure~\ref{figure1}, in black (pure rotational lines), and grey (rovibrational lines, when available). The data were corrected for extinction, using A$_{\rm V}=2$ from \citet{Giannini04}. The inferred rotational temperatures are indicated in black, and the rovibrational temperatures, in grey. The best-fitting model results are also shown in each case, in blue and/or red symbols. The shock parameters are given in each panel in the corresponding colour.}
         \label{Figure2}
   \end{figure*}

The data were corrected for extinction, adopting the upper limit of the visual extinction from \citet{Giannini04}, $A_{\rm V} = 2$, and using the interstellar extinction law of \citet{Rieke85}. We note that the effect of the interstellar extinction on the observational points is minor. The excitation diagrams resulting from the observations are plotted in Figure~\ref{Figure2}. The rotational and rovibrational temperatures derived from these observations are also indicated on the plots in the same colour code, along with the corresponding line. Again, the reddening correction has only minor significance on the temperature values. As anticipated, the rovibrational transitions are found to trace warmer gas than the pure rotational transitions (see, for example, \citealt{Lebourlot02, Giannini06}, and G08b). 

\subsection{Grid of models}
\label{sub:gom}

\begin{table}
\caption{Input parameters of our grid of stationary C- and non-stationary CJ-type shock models.}             
\label{table2}      
\centering                          
\begin{tabular}{c c c c}        
\hline            
type & $n_{\rm H}$ (cm$^{-3}$) & b & $\varv_{\rm s}$ (km s$^{-1}$)  \\    
\hline \hline  
C & 10$^4$ & $\in$ [0.45 ; 2] & \tiny{10,15,20,22,25,28,30,32,35,55} \\
C & 10$^5$ & $\in$ [0.30 ; 2] & \tiny{10,12,15,18,20,40} \\
C & 10$^6$ & 1 & \tiny{10,12,15,18,20,22,25,27,30,32} \\
\hline                
CJ & 10$^4$ & $\in$ [0.45 ; 2] & \tiny{10,12,15,20,22, 25,28,30,32,35} \\
CJ & 10$^5$ & $\in$ [0.30 ; 2] & \tiny{10,12,15,18,20} \\
\hline
\end{tabular}
\end{table}

The shock code that we use to simulate the propagation of shock waves in the interstellar medium is described in G08b. It generates one-dimensional stationary models of both continuous or discontinuous kind (respectively, C- or J- type shocks), and also approximations to time-dependent solutions, in the form of a shock wave possessing the characteristics of both C- and J-type models (see \citealt{Chieze98,Lesaffre041,Lesaffre042}). The shock code solves the differential equations that determine the evolution of a set of dynamical and chemical variables: the neutral and ionized fluid temperatures, velocities and densities; the fractional abundances of over 125 species, linked by more than 1000 chemical reactions. The populations of the first 150 levels of molecular hydrogen are calculated within the shock code, based on the treatment of  \citet{Lebourlot02}, with the only difference being that we adopted the most recent H--H$_2$ collisional rate coefficients (\citealt{Wrathmall07}). Finally, grain-related processes are also included in the code, following \citet{Flower031}. 

The parameters that we aim to constrain are the inputs of the models, namely:
\begin{itemize}
\item ambient-medium-related quantities: pre-shock density $n_{\rm H}$ and magnetic-field parameter $b$, defined by B($\mu$G)~=~$b~\times~\sqrt{n_{\rm H}~({\rm cm}^{-3})} $;
\item shock-related quantities: shock velocity $\varv_{\rm s}$, and age in the case of a non-stationary solution;
\item chemistry-related quantities, such as the initial value of the ortho-para ratio for molecular hydrogen, or the initial distribution of silicon-bearing material in the grains.
\end{itemize}

As a first step, we made use of a grid of stationary C-type shock models with the parameters specified in Table~\ref{table2}. We should mention that the parameter coverage is not complete, in the sense that not all velocities are present in our grid for all values of the magnetic-field parameter, $b$. In fact, the velocity of C-type shocks must remain below a critical value that depends mainly on the pre-shock density and magnetic-field parameter (\citealt{Lebourlot02,Flower031}), which explains the decrease of the maximum shock velocity with the pre-shock density in our grid. Following the method presented in G08b (Section~4.1), this set of C-type shock models enabled us to restrict the range of the search in the parameter space for the CJ-type shock models. We then computed a grid of non-stationary shock models around a first estimate of the shock age, making sure that the range of ages was sufficient to include any model likely to fit the H$_2$ observational data. The grid of CJ-type models covers the combination of parameters given in Table~\ref{table2}, for ages ranging from a few tens to a few thousands of years.

For each of the above models, we obtained an excitation diagram that can be compared with observations.

\subsection{Comparisons and initial constraints}
\label{sub:cafc}

We describe here the results of the comparisons between observations and our models. A few preliminary remarks about our results are applicable to all the knots and regions analysed. First, the filling factor was considered to be equal to 1 for all transitions; this yields H$_2$ knot sizes consistent with the thickness of the H$_2$ emitting layer of our models. Secondly, under this assumption, we find that, in all cases, only non-stationary shock models reproduce satisfactorily both the pure rotational and rovibrational (when available) parts of the H$_2$ excitation diagram, a result that has already been established in similar studies of other objects (e.g. HH54, \citealt{Giannini06}, or L1157-B1, see G08b). Finally, the departure of the observational points from a straight line in the purely rotational parts of the excitation diagram (\lq saw-tooth' pattern, which is weak in our cases) is indicative of deviations of the ortho-to-para ratio from the high-temperature LTE value of 3.0. Such a pattern can be reproduced in the models by slightly modifying the value of the initial ortho-to-para ratio; but fine-tuning this parameter is not a major consideration in the context of the present study.

We give prior focus to the Herbig-Haro objects, for which we can benefit from H$_2$ excitation diagrams combining pure rotational and rovibrational transitions, and to the so-called \lq SiO knot', for which future observations of SiO rotational transitions are expected to provide further constraints on our shock models. Tables~\ref{tablea1} and \ref{tablea2} present a summary of the restrictions that comparisons between observational and computed excitation diagrams place on the shock model parameters. In these tables, the values of the pre-shock density, magnetic-field parameter and shock velocity are indicated for our best models, as well as the age range for which the excitation diagrams are in satisfactory lagreement with the observations. In the case of the Herbig-Haro objects HH320A\slash B and HH321A\slash B, the best-fitting models are non-stationary, and the pre-shock density is constrained to 10$^4$~cm$^{-3}$, which is in agreement with the average density of the whole globule (9$\times$10$^{3}$~cm$^{-3}$), as determined by \citet{Bourke97}. Shock velocities in the range 15--35~km~s$^{-1}$ yield good fits to the observations, depending on the associated magnetic-field parameter and age. We note that the shock ages that we obtain are of the same order of magnitude as the observational dynamical ages. The excitation conditions are remarkably similar in both sources, in spite of their location on a different lobe of the outflow, and the probable mixing of the two outflows from IRS1 and IRS2. In the case of HH320A\slash B, our findings are in agreement with those of \citet{Giannini04} in terms of pre-shock density, magnetic field, and shock type and age, but our value of the shock velocity is smaller than theirs. Several factors could account for this difference, which might not be a real discrepancy, given the degeneracy of our fitting models. The first factor is that their value lies slightly outside the range of our parameter search. In addition, we studied HH320A\slash B, rather than HH320A. Furthermore, we added the pure rotational transitions of H$_2$ to the analysis. Finally, our shock model incorporates an update of the H--H$_2$ collisional rate coefficients, following \citet{Wrathmall07}.

The parameters of the SiO knot are not as well constrained, as only the pure rotational observations are available to build the H$_2$ excitation diagram. Similarly to the HH objects, the best fitting models are of non-stationary shocks with velocities in the range of 15--35~km~s$^{-1}$ for a pre-shock density of 10$^4$~cm$^{-3}$ (with ages dependent on the value of the magnetic-field parameter). Additionally, a few models with a pre-shock density of 10$^5$~cm$^{-3}$ are found to fit the observations equally satisfactorily, with even more moderate shock velocities (10--20 km~s$^{-1}$) and lower shock ages (which again depend on the value of the magnetic-field parameter). Examples of the comparisons between observations and models are shown in the three upper panels of Figure~\ref{Figure2}, for the cases of the SiO knot, HH320A\slash B and HH321A\slash B, respectively. In each panel, the parameters corresponding to the selected shock model are also indicated.

For the three other points studied(\lq knot 5', \lq knot 6', and \lq knot 8'), once again the excitation diagrams do not constrain the parameters as tightly as in the HH regions, since only the pure rotational transitions of H$_2$ can be used. The number of models that fit these data is large, but we show some examples of the best fits that we could derive in the lower panels of Figure~\ref{Figure2}. We note that the modelling work conducted for these positions also suggests similar excitation conditions, regardless of their association with one lobe of the outflow or the other.

\section{SiO observations and modelling}
\label{sec:SiOoam}

\subsection{Observations}
\label{sub:SiOo}

The observations were carried out mainly in two observational runs, in August and October 2009, with the Atacama Pathfinder EXperiment (APEX\footnote{This publication is partly based on data acquired with the Atacama Pathfinder EXperiment (APEX). APEX is a collaboration between the Max-Planck-Institut f\"ur Radioastronomie, the European Southern Observatory, and the Onsala Space Observatory.}, \citealt{Guesten06}). We used the facility's own APEX--1 and APEX--2 receivers (see respectively \citealt{Vassilev08}, \citealt{Risacher06}) to observe respectively the SiO (5--4) and (6--5) transitions, and the SiO (8--7) transition, in combination with the MPIfR Fast Fourier Transform Spectrometer backend (FFTS, \citealt{Klein06}). Focus was checked at the beginning of each observing session, and after sunrise on Mars. Line pointing was locally checked on 07454--7112 or IRAS 15194--5115. The pointing accuracy was found to be of the order of 5$''$ rms. Table~\ref{table3} contains the main characteristics of the telescope for the observed transitions:~ beam sizes, forward and beam efficiencies, typical system temperatures, and the spectral resolution that was used for each transition. The observations were performed in position-switching mode using the APECS software (\citealt{Muders06}), with a reference position at offset (-600$''$,600$''$) from the central position indicated in Table~\ref{table1}. The data were reduced with the CLASS software (see http://www.iram.fr/IRAMFR/GILDAS). 

\begin{table}
\caption{Observed lines, frequencies, and corresponding telescope parameters:~beam and forward efficiencies, beam sizes, typical system temperatures, and spectral resolution.}             
\label{table3}      
\centering                          
\begin{tabular}{c  c  c  c }        
\hline           
line & SiO (5--4) & SiO (6--5) & SiO (8--7) \\
\hline \hline
$\nu$ (GHz) & 217.105 & 260.518 & 347.331 \\
\hline
$B_{\rm eff}$ & 0.75 & 0.75 & 0.73 \\
$F_{ \rm eff}$ & 0.95 & 0.95 & 0.95 \\
\hline
beam ($''$) & 28.7 & 24.0 & 18.0\\
$T_{\rm sys}$ (K) & 155--175 & 270--300 & 275--335 \\
$\Delta$v (km s$^{-1}$) & 2.02 & 1.69 & 1.26 \\    
\hline                                  
\end{tabular}
\end{table}

\subsection{Results}
\label{sub:SiOr}

The observations resulted in one SiO (5--4) map of the inner, upper (red) lobe, as well as in SiO (6--5) and (8--7) spectra, obtained by pointing the telescope at designated positions inside this lobe. The absolute coordinates of the (0$''$,0$''$) position are given in Table~\ref{table1}. 

\begin{figure}
\centering
\includegraphics[height=9cm,angle=-90]{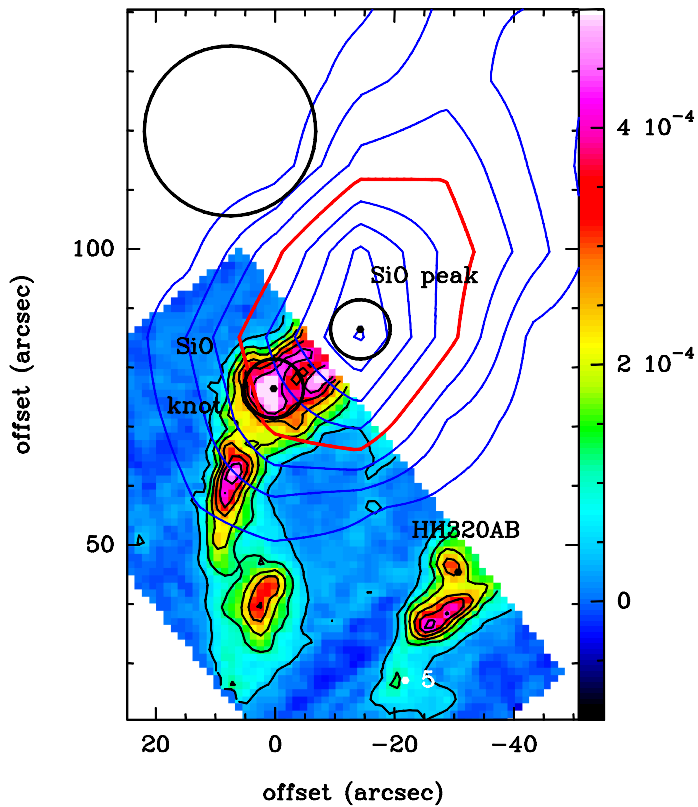}
\caption{Upper internal part of the BHR71 outflow. The background map is the same as in Figure~\ref{figure1}. The SiO (5--4) contours, which are overlaid in a dark blue colour, correspond to levels of 6 to 30~$\sigma$, in steps of 3~$\sigma$, for an integration between $-$10 and 60 km s$^{-1}$. The half-maximum, 15~$\sigma$ contour is displayed in red. The associated beam size is indicated in the upper left corner. Our SiO (5--4) coverage is actually wider than the field displayed on this Figure.}
\label{figure3}
\end{figure}

The map of the SiO~(5--4) line covers a roughly rectangular region, extending from the lower left offset (35$''$,25$''$) to the upper right (-60$''$,170$''$), and was performed in raster mode, spending 30~seconds on each point. The beam sampling was 10$''$, which is less than the half beam size of 14.35$''$, meaning the map is fully Nyquist sampled. The result is plotted in Figure~\ref{figure3}: the SiO emission, integrated between $-$10 and 60~km~s$^{-1}$, is displayed in dark blue contours overlaid on the H$_2$~0--0~S(5) map already shown in Figure~\ref{figure1}. The half-maximum signal contour is shown in red. The r.m.s. of the SiO~(5--4) map was found to be 0.26 K~km~s$^{-1}$ at the resolution 2.02~km~s$^{-1}$, and the contours correspond to levels of 6 to 30~$\sigma$, in steps of 3~$\sigma$. The contours show that the SiO~(5--4) emission peaks outside of the \textit{Spitzer} spectral line maps, at a position that will be referred to as the \lq SiO peak', with offset (-15$''$, 87$''$). In order to perform a simultaneous study of the H$_2$ and the SiO emission, the point to be studied must be located inside the \textit{Spitzer} map:~we chose the point of maximum emission of H$_2$, at an offset of (0$''$, 77$''$), to which we refer as the \lq SiO knot'. We consider the shift between the APEX SiO~(5--4) and \textit{Spitzer} H$_2$~0--0~S(5) emission peaks to be real, as it is far too large to be attributable to pointing errors. Indeed, the distance between the south-eastern and the north-western tips of the half-maximum SiO emission (roughly of respective coordinates (0$''$,70$''$) and ($-$30$''$,110$''$), see Figure~\ref{figure3}) is about 50$''$, which amounts to 40$''$ once deconvolved from the SiO~(5--4) telescope beam size. Consequently, we consider that, in this direction, which coincides with the line between our SiO knot and peak (see Figure~\ref{figure3}), the emitting region is extended enough to allow us to consider that the signal observed at the position of the SiO knot is free of any residual signal from the SiO peak. Moreover, from the extension of the \textit{Spitzer} maps, it is impossible to exclude the presence of another, potentially stronger peak of H$_2$ emission, coinciding with the SiO peak. In order to check, we overlaid our map on the four (way more extended) channel maps of IRAC, especially the fourth, at 8~microns, whose broadband intensity is dominated by the 0--0~S(4) and S(5) transitions of H$_2$ in this region (\citealt{Neufeld09}). We found that these overlays fail to confirm the existence of an H$_2$ peak coinciding with that of the SiO~(5--4) emission.

\begin{figure}
\centering
\includegraphics[height=9cm,angle=-90]{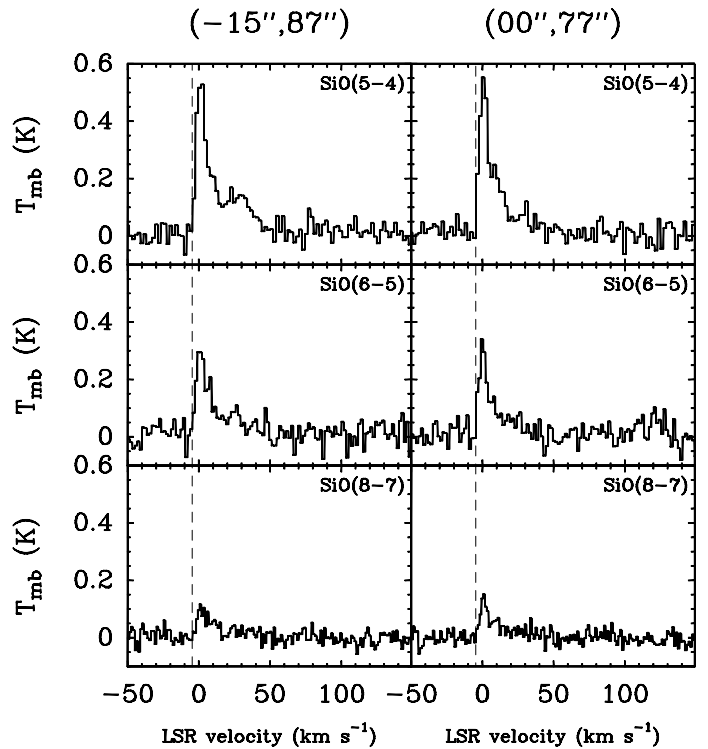}
\caption{The spectra obtained in the three observed SiO transitions at the positions of the SiO peak and knot, of respective offsets (-~15$''$,87$''$) and (0$''$, 77$''$). Only the SiO knot is considered in our modelling. For each spectrum, we adopted the SiO (5--4) spatial resolution. The ambient cloud velocity, -4.5 km s$^{-1}$ is indicated by the vertical dashed lines.}       
\label{figure4}
\end{figure}

The spectra that we obtained of the SiO (5--4) transition at both the SiO peak and knot are presented in Figure~\ref{figure4}, along with the spectra of the other two SiO transitions. In the case of the SiO (5--4) transition, the original beam size was 28.7$''$. In order to derive the spectra at the two selected positions as exactly as possible, we resampled the data with a spatial resolution of 5$''$ and then averaged the four spectra closest to the two positions to obtain the results that are presented on Figure~\ref{figure4}. In the case of the SiO (6--5) and (8--7) transitions, in addition to this spatial resolution problem, we had also to convolve the data to the SiO (5--4) beam size in order to render the whole set of transitions consistent. After this preliminary convolution, we obtained the spectra displayed in Figure~\ref{figure4}. Our mini-maps around the two positions were extended enough to cover the larger beam of the SiO (5--4) observations.

The shape of the line profiles is typical of bipolar outflows (similar, for instance, to those acquired around L1157: see \citealt{Nisini07}), peaking near the cloud velocity and with the main signature being the line wings extending to about 50~km~s$^{-1}$, most visible in the SiO(5--4) line profile of the peak. Such wings seem to be present, though narrower, at higher frequencies. However, it is then more difficult to estimate their width, as the signal-to-noise ratio decreases as the observed frequency increases. The observed SiO probably arises from a complex mixture of shocks with different velocity projections on the line of sight. The velocity range of our shock models does not exceed the terminal velocity of 50~km~s$^{-1}$. On the SiO (5--4) and (6--5) spectra associated to the SiO peak, a velocity bullet seems to arise between 20 and 50~km~s$^{-1}$. It could arise from a process that alters the SiO emission within the region of the shock considered, such as an abundance effect, or from the presence of another shock, with a different velocity, within the telescope beam. These possibilities are discussed in Subsection~\ref{sub:mr} but cannot be thoroughly investigated, given the large beam sizes associated with these transitions; they call for a more detailed study, with higher spatial resolution.

The last step was to derive integrated intensities from these spectra. The results for the SiO knot, which is considered in our subsequent analysis, are shown in Table~\ref{table4}:~integrated intensities and associated r.m.s. are given. The velocity interval over which the integration was made was [$-$10, 60]~km~s$^{-1}$ for all three transitions. Again, the values inferred are consistent with results for other outflows, such as L1448 or L1157 (\citealt{Nisini07}), for which the central source is also a young Class 0 proto-star with a similar bolometric luminosity (respectively, around 13.5, 11, and 1.5--3.5~L$_\odot$ for BHR71, L1157, and L1448; see \citealt{Neufeld09}). In particular, the line profiles we observe for SiO (5--4), (6--5), and SiO(8--7) are remarkably similar to those of L1157-B1, peaking near the cloud velocity. The integrated intensities we inferred are also very close to those compiled for each knot of the L1157 bipolar outflow (\citealt{Nisini07}).

\begin{table}
\caption{Integrated intensities and r.m.s. for the three observed SiO transitions at the position of the SiO knot, of relative coordinates (0$''$, 77$''$).}             
\label{table4}      
\centering                          
\begin{tabular}{c c c c}        
\hline            
line & SiO (5--4) & SiO (6--5) & SiO (8--7) \\    
\hline \hline          
$\int T_{\rm{MB}} \Delta \varv$ (K km s$^{-1}$) & 7.7 & 4.1 & 1.8 \\      
$\sigma_{\int T_{\rm{MB}} \Delta \varv}$ (K km s$^{-1}$) & 0.4 & 0.4 & 0.2 \\
\hline                                  
\end{tabular}
\end{table}

\subsection{Modelling results}
\label{sub:mr}

We derive emission-related quantities by means of a radiative transfer code used in combination with the shock models presented above. The radiative transfer code is based on the LVG approximation, and has been described and used in G08a and G08b. At each point of the shock model, it uses the computed parameters of relevance, such as the temperature, density, velocity gradient, and collision-partner density, to solve the equations of statistical equilibrium and calculate emission-related quantities, such as the level populations, optical depths, and local emissivities. The set of equations is solved by means of a simple lambda-iteration, that stops when the relative difference between each of the level populations computed in two successive iterations drops below a convergence value of 10$^{-4}$.  The set of collisional rate coefficients for SiO with H$_2$ as a collision partner is simply scaled by the square root of the reduced mass ratio from the SiO--He rate coefficients calculated by \citet{Dayou06}. The LVG code also provides the integrated intensity over the whole shock model for all transitions considered.

\begin{figure}
\centering
\includegraphics[width=6cm]{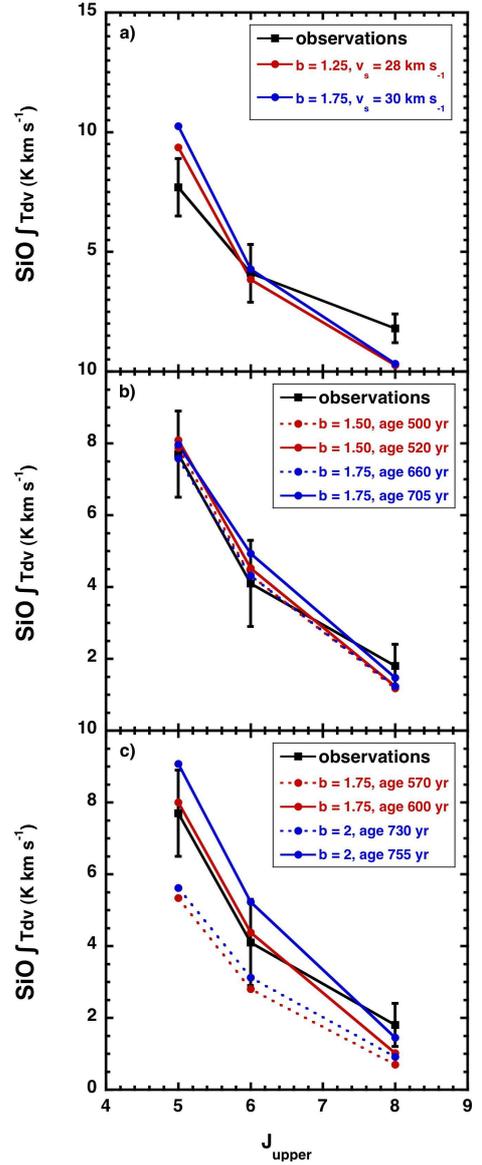}
\caption{Examples of fits of the integrated intensities of our three transitions of SiO. On the $y$-axis is the integrated intensity (K~km~s$^{-1}$), and on the $x$-axis is the rotational quantum number of the upper level of the transition. The observations are the black squares linked by the thick black lines, on all three panels. The 3$\sigma$ errorbars are shown in the Figure. The model results are displayed in colour (blue and red), with the indication of the specific shock parameters on each panel. (a) C-type shock models: best fit shock parameters $n_{\rm H} = 10^4$~cm$^{-3}$, no SiO initially in the grain mantles; (b) CJ-type shock models: best fit shock parameters $n_{\rm H} = 10^4$~cm$^{-3}$ and $\varv_{\rm s} = 28$~km~s$^{-1}$, with 1\% of elemental silicon initially in the grain mantles, as SiO; (c) CJ-type shock models: best fit shock parameters $n_{\rm H} = 10^4$~cm$^{-3}$ and $\varv_{\rm s} = 30$~km~s$^{-1}$, with 10\% of the silicon initially in the grain mantles as SiO.}         
\label{figure5}
\end{figure}

The silicon chemistry is extensively discussed in G08a and G08b. Regarding the initial distribution of silicon-bearing material, G08a made the assumption that Si is initially present exclusively in the grain cores, in the form of silicates, with a fractional abundance of $3.35 \times 10^{-5}$ relative to total H. G08b introduced an alternative scenario, in which the Si is partly in the grain mantles, in the form of SiO. Such an assumption was found to be necessary to fit both the SiO and H$_2$ line intensities by means of a single shock wave model. Furthermore, a simultaneous fit can be achieved only by means of a non-stationary, CJ-type shock. When computing CJ-type models in the present study, we adopt one of three values of the fraction of elemental silicon initially in the form of SiO in the grain mantles, namely 0.00, 0.01, or 0.10.

In the first case, when no SiO is initially present in the grain mantles, the gas-phase silicon is produced by the erosion of the grain cores only, which occurs, in the C- and CJ-shock models, where the predominantly charged grains collide with the neutral particles at the ion-neutral drift speed (ambipolar diffusion). Three reaction paths then lead to the formation of SiO in the gas phase, namely Si$^+$(OH,H)SiO$^+$(H$_2$,H)HSiO$^+$(e,H)SiO, Si(OH,H)SiO, and Si(O$_2$,O)SiO; the last two reactions predominate. Following G08a, we adopted the rate coefficient that was experimentally measured for the Si+O$_2$ reaction by \citet{Lepicard01} for both Si+O$_2$ and Si+OH. In the second case, in which SiO is initially present in the grain mantles, their rapid erosion by sputtering in the shock wave generates SiO directly in the gas phase (see also the discussion in G08b).  

O$_2$ is a key reaction partner in the process of formation of SiO in the gas phase, and so its abundance must be modelled to the greatest possible accuracy. Furthermore, the question of the pre-shock abundance of O$_2$ in the gas phase must be considered carefully. In all cases, as in G08a and G08b, we used an initial gas-phase abundance of $n(\rm{O_2})/n_{\rm H} = 1.0 \times 10^{-7}$, consistent with the O$_2$ observations of the Odin satellite (see \citealt{Pagani03,Larsson07}). On the other hand, the initial fractional abundance of O$_2$ in chemical equilibrium is $1.0 \times 10^{-5}$; we allocated the difference between these two values (i.e. most of the initial O$_2$) to the grain mantles, consistent with G08a and G08b. Given that the O$_2$ in the mantles is released rapidly into the gas phase, by sputtering processes induced by the shock wave, this scenario does not differ significantly from that in which all of the O$_2$ is assumed to be present initially in the gas phase (the chemical equilibrium situation). An alternative hypothesis, in which the same fraction of the corresponding elemental oxygen is assumed to be in the form of water ice, rather than O$_2$ ice, was considered by G08a and found not to modify significantly the results of the models. Indeed, in this case, the rapid erosion of the grain mantles during the passage of the shock wave releases H$_2$O into the gas phase, and the subsequent reaction of H$_2$O with H produces OH, which then reacts with Si to form SiO. A third scenario was put to the test in the present study: placing the elemental oxygen in the grain mantles in the form of CO$_2$ ice. In this case, CO$_2$ is released quickly into the gas phase during the passage of the shock wave, but no chemical reactions convert it to O$_2$ or OH. When SiO is present in the grain mantles, direct sputtering is the dominant source of the gas-phase SiO. However, when SiO is absent from the grain mantles, erosion of the silicon-bearing grain cores is the mechanism that releases silicon into the gas phase, and the deficit in O$_2$ or OH leads to lower abundances of gas-phase SiO and weaker SiO line emission from low-velocity shocks. In high-velocity shock models, or at higher pre-shock densities, O$_2$ and OH are destroyed in the gas-phase, through the O$_2$(H,O)OH and OH(H,O)H$_2$ sequence, and so the initial composition of the grain mantles is less significant.

If one ignores the observations of H$_2$, it is possible to fit the observations of SiO by means of  C-type shock models, without any SiO initially in the grain mantles; cf. G08a. In this case, we find that shock models with a pre-shock density of 10$^4$~cm$^{-3}$ provide better fits than those with pre-shock densities of 10$^5$ or 10$^6$~cm$^{-3}$. The typical shock velocities are from 25 to 30 km s$^{-1}$, with values of the magnetic-field parameter from 0.45 to 2. An example of such a fit can be seen in panel (a) of Figure~\ref{figure5}. We note that, in this case, the assumption that most of the O$_2$ is initially in the form of CO$_2$ ice on the grain mantles leads to insufficient levels of SiO emission, as compared with the observations. However, our aim is to model simultaneously the emission of SiO \textit{and} H$_2$, assuming their emitting regions coincide, to within the spatial resolution of the observations. Accordingly, we calculate the SiO spectrum for the shock models that already fit the H$_2$ lines, namely the non-stationary shock models presented in Subsection~\ref{sub:cafc}. Given the the additional flexibility afforded to these models by the initial fractional abundance of SiO in the grain mantles, we found that most of the CJ-type models that already fit the H$_2$ data can reproduce the SiO emission satisfactorily. The results are presented in Figure~\ref{figure5}, where panels b) and c) correspond to our best fitting models with, respectively, 1\% and 10\% SiO initially in the mantles. We find that no model in which SiO is absent from the grain mantles is able to reproduce the intensities of the SiO lines simultaneously with those of H$_2$, thereby confirming a conclusion of G08b and rendering considerations of the initial chemical form of oxygen irrelevant.

We wish to emphasize that a single, plane-parallel CJ-type shock wave model is unlikely to provide a completely satisfactory simulation of the molecular line emission from an object as complex as BHR71. In particular, a combination of C-type shock models with appropriate parameters and filling factors might also be able to reproduce the H$_2$ and SiO line intensities and address the problem of fitting the SiO line profiles. The latter could be strongly affected by geometrical effects, associated with the projection of multiple shock velocities on to the line of sight. However, such a study introduces additional free parameters that are insufficiently constrained by the currently available observations. Furthermore, our model does not take into account the consequences of grain-grain collisions, whose inclusion (following \citealt{Guillet07, Guillet09, Guillet11}) would have major implications for computation times. In view of these limitations, geometrical and physical, it is not feasible currently to investigate the nature of the bullet that we observe in the SiO(5--4) line profile.

\section{Water emission predictions}
\label{sec:wep}

\subsection{Water modelling}
\label{sub:wm}

The abundance of water at each step of the calculation is the result mainly of two formation pathways. One is the O(H$_2$,H)OH(H$_2$,H)H$_2$O sequence, which takes place in the gas phase and becomes significant when the temperature exceeds a few hundred kelvin (e.g. \citealt{Charnley97,Atkinson04}). The other is the sputtering of water ice that formed in the mantles from O and H atoms, which were adsorbed on to the grains in the parent dark cloud (see for example \citealt{Tielens82}). In the latter case, the sputtering process is analogous to that considered for SiO when present in the grain mantles.

The modelling of water emission was performed in a similar manner to that of SiO. Again, we used an external radiative transfer module based on the LVG approximation; the molecular data were taken from the LAMDA database (http://www.strw.leidenuniv.nl/$\sim$moldata/). Regarding the rate coefficients for excitation of H$_2$O by He, we used the data of \citet{Green93}. For collisions of H$_2$O with H$_2$, we adopted the same approach as FP10 and used the CTMC results of \citet{Faure07}, rather than the more precise but less complete (in terms of their temperature range) quantal calculations of  \citet{Dubernet02}, \citet{Grosjean03}, and \citet{Dubernet09}. As no data exist for the excitation of H$_2$O in collisions with H, we used the same rate coefficients as for excitation by ortho-H$_2$, following FP10. 

The more complex structure of H$_2$O, as compared to SiO, results in some changes being required to the computational method in the radiative transfer module. First, the code must be  modified to take account of the more complicated radiative selection rules. Second, in the more complex case of water as compared to SiO, a value of 10$^{-10}$ is required for the convergence parameter. Finally, we found it necessary to adapt the expression that is used for the escape probability. For a plane-parallel geometry, such as is assumed in our shock models, the escape probability is

\begin{equation}
\label{eq:betaiso}
\beta_\perp = \frac{1 - e^{-\tau_\perp}}{\tau_\perp}.
\end{equation}
where $\tau_\perp$ is the LVG optical depth in the $z$-direction, normal to the shock front. On the other hand, \citet{Neufeld93} proposed the following approximation to the angle-averaged escape probability,
\begin{equation}
\label{eq:betaplane}
\bar\beta_{\rm av} = \frac{1}{1 + 3\tau_\perp},
\end{equation}
In the case of water, we found that the use of the second formula gave rise to numerical instabilities and longer computation times. Consequently, we adopted the first formula, both when solving the equations of statistical equilibrium and when evaluating the local line emissivity at each point in the shock wave. This approach is consistent with that of FP10 and enables us to make direct comparisons with results obtained using their method of treating radiative line transfer in shock waves. The implications of the different forms of the escape probability for the intensities of the rotational transitions of SiO are discussed in Appendix~A of G08a. We recall that the geometry of the shock wave is not well known.

Before comparing with the method of FP10, we benchmarked our LVG code against the RADEX code (\citealt{Vandertak07}), at four representative points, covering different conditions of temperature, density, and velocity gradient, namely (respectively in K, cm$^{-3}$, and s$^{-1}$): ($25$, $10^4$, $2\times10^{-13}$), ($50$, $10^5$, $10^{-13}$), ($300$, $10^4$, $10^{-12}$), and ($467$, $3.71\times10^5$, $9\times10^{-11}$). For the purpose of this comparison, we adopted the same escape probability formula as in the RADEX code. For ortho-H$_2$O, the computed excitation temperatures showed no mean relative discrepancy above 6.3\% for the ten levels emitting the strongest lines at each of the four points. In the case of para-H$_2$O, the relative discrepancy was occasionally larger, reaching 43.9\%. Although it remains unclear why the discrepancies were larger for para- than for ortho-H$_2$O, the overall level of agreement with the RADEX code was reassuring.

\subsection{Steady-state versus statistical-equilibrium}
\label{sub:ssse}
   
      \begin{figure*}
   \centering
   \includegraphics{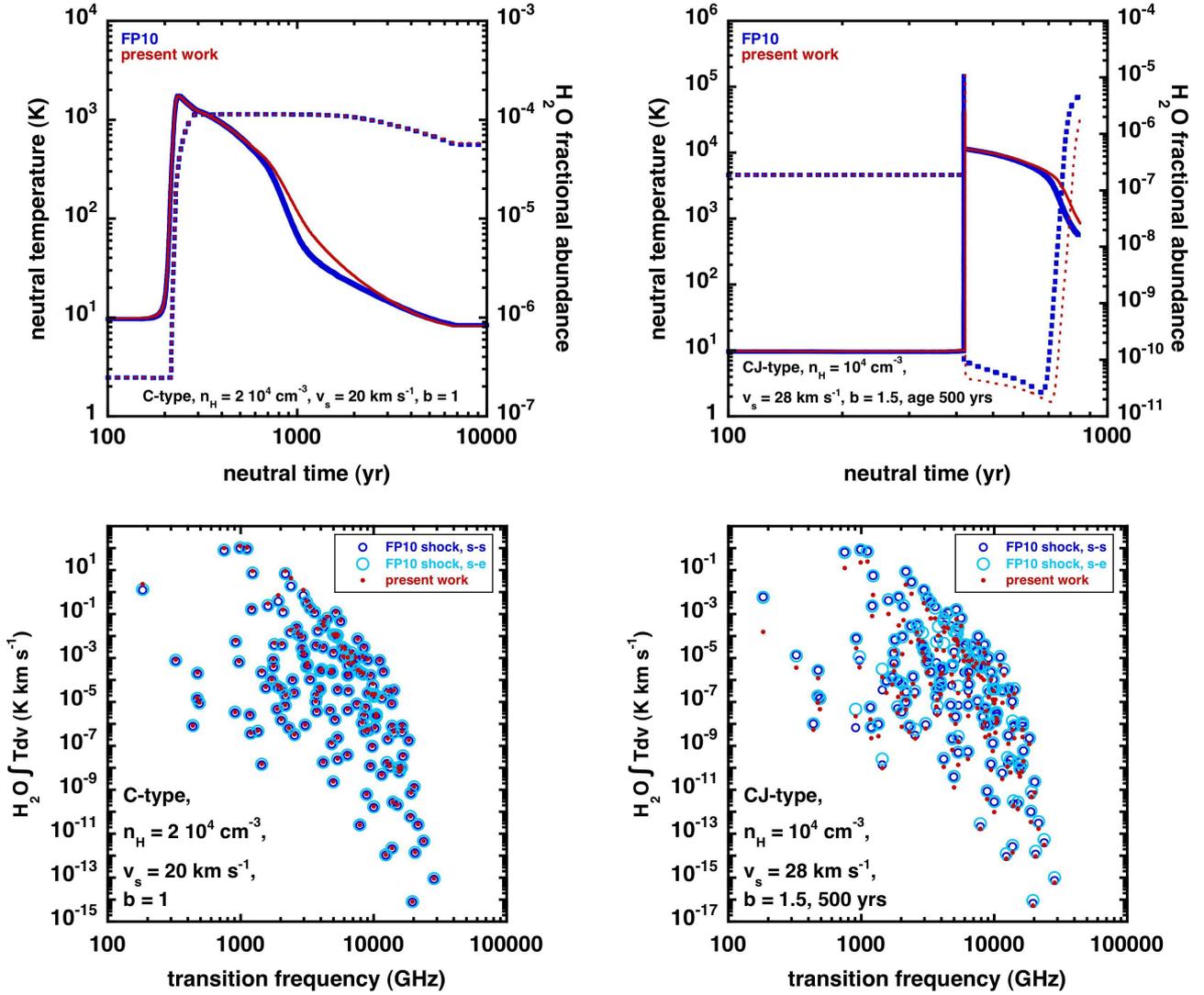}
      \caption{Upper panels:~comparison between profiles obtained when the computation of the water line emission is internal to the shock code (steady-state approximation: blue curves) or external to the shock code (statistical equilibrium approximation: red curves);~the temperature of the neutral fluid (continuous line) and the fractional abundance of water (dashed lines) are shown for our two reference, C-type (left-hand panel), and CJ-type (right-hand panel) shock models. In both cases, the shock parameters are displayed in the corresponding panels. Lower panels:~integrated line intensities for all para-water transitions;~steady-state approximation to the computation of the water emission (dark blue open circles); same shock model, but with the statistical-equilibrium approximation to the computation of the water emission (large light-blue open circles); present shock model, with the statistical-equilibrium assumption (small red filled circles).}
         \label{Figure6}
   \end{figure*}

       \begin{figure*}
   \centering
   \includegraphics{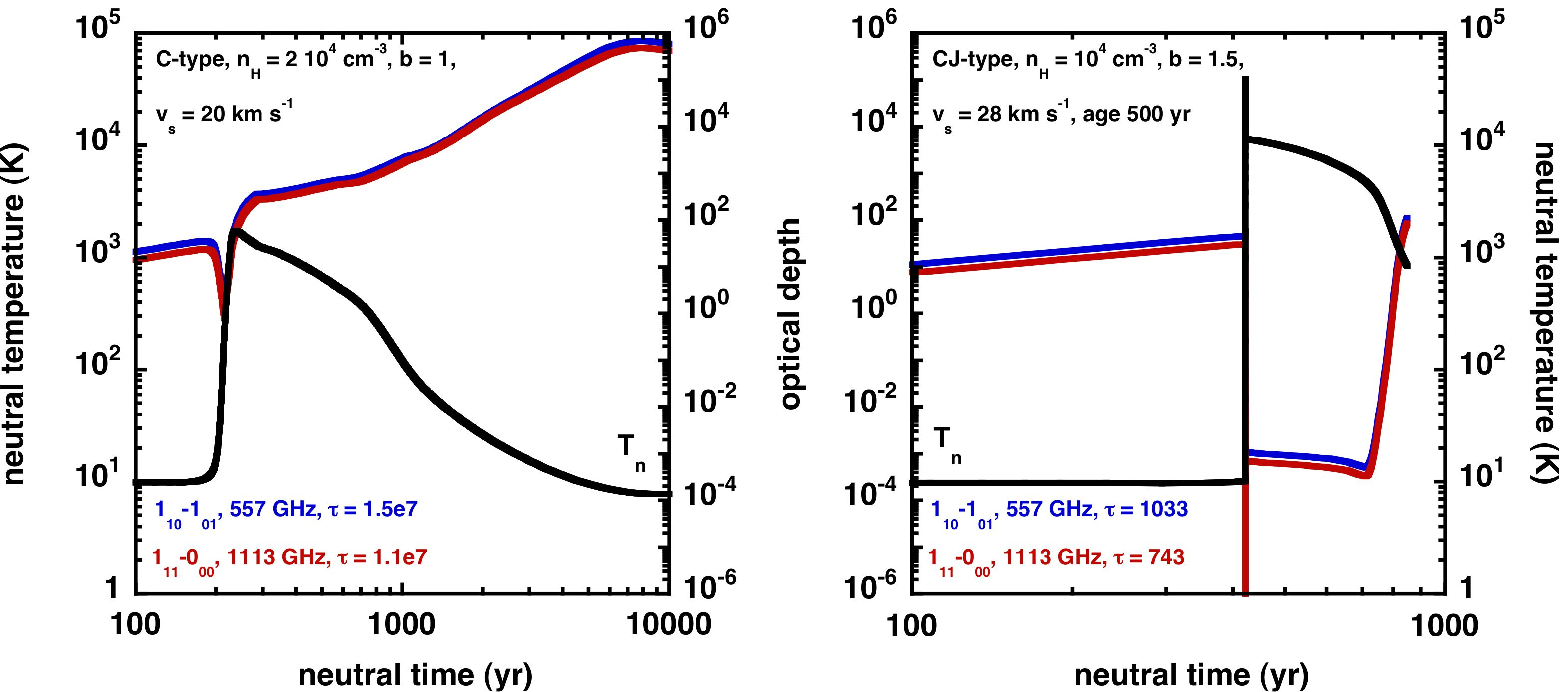}
      \caption{Evolution of the computed local optical depths of two transitions as a function of the flow time of the neutral fluid in our reference models, the stationary C-type (left panel) and non-stationary CJ-type (right panel). The o-H$_2$O 1$_{\rm{10}}$--1$_{\rm{01}}$ (at 557~GHz) and p-H$_2$O 1$_{\rm{11}}$--0$_{\rm{00}}$ (at 1113~GHz) lines are considered, and their optical depths are shown in blue and red, respectively. The corresponding total optical depths are indicated in the same colour code.  The neutral temperature profile shown in black. The shock parameters are specified in each panel.}
         \label{Figure7}
   \end{figure*}  

In this section, we compare the results of the present calculations with those obtained using the approach of FP10. We present the intensities of all ortho- and para-transitions, integrated over the whole extent of the shock wave. We compared one stationary C-shock model, with input parameters $n_{\rm H}~=~2\times$10$^4$~cm$^{-3}$, $\varv_{\rm s}~=~20$~km~s$^{-1}$, and b = 1, and one non-stationary CJ-shock model, with input parameters $n_{\rm H}~=$~10$^4$~cm$^{-3}$, $\varv_{\rm s}~=~28$~km~s$^{-1}$, b = 1.5, and age 500~yrs, which fits both the H$_2$ and SiO observations (see Figure~\ref{figure5}, panel b). The principal difference between the two versions of the shock code is that FP10 included the population densities of the levels of water as variables of the model, and the differential equations describing the evolution of the level populations were solved simultaneously with those for the other dynamical and chemical variables, in the steady-state approximation. Thus, the water line emissivities were computed within the shock code, at each point of the shock profile, and could be integrated internally. In the present approach, the level population densities (in statistical-equilibrium) and the line intensities are calculated in a separate LVG module, to which the shock profiles are provided as input data.

In addition to the different approximations used when calculating the population densities of the water levels (steady-state\footnote {$\partial/\partial \it{t}$ = 0 and hence $\rm{d} / \rm{d} \it{t} = \it{\varv} \rm{d} / \rm{d} \it{z}$} and statistical-equilibrium\footnote{$\rm{d} / \rm{d} \it{t} \equiv 0$}), there are two further, potentially significant differences between these two sets of calculations. In the steady-state case, the radiative cooling function that is calculated at each point includes the contribution of the water emission, determined from the level population densities and the line escape probabilities. This contribution has an effect on the temperature profile in regions where H$_2$O is a significant coolant. In the statistical-equilibirum assumption, the contribution of water emission to the cooling of the medium is approximated by means of the tables provided by \citet{Neufeld93}. In addition, the effect of adsorption of gas-phase species on the radius of the grains is treated differently in the two versions of the shock code; this affects the gas--grain interactions in the tail of the cooling flow, where adsorption takes place. As a consequence, the coupling of the gas to the grain temperature, as the post-shock region is approached, occurs differently in the two versions of the shock code.

In both the C-type and CJ-type cases, we have three sets of LVG calculations. We denote by `FP10'  the results of the steady-state (s-s) shock code. Additionally, the dynamical and chemical profiles generated by either version of the shock code may be combined with a separate,  statistical-equilibrium (s-e) treatment of the water level populations. The upper panels of Figure~\ref{Figure6} compares the neutral temperature and water fractional abundance profiles from the two versions of the shock code. In the case of the C-type model, differences are noticeable only in the cooling flow, where water is a significant coolant (see top right panel of Figure~2). Nonetheless, the differences in the thermal profiles are modest, and the water fractional abundances in the two versions remain indistinguishable. In the case of the CJ-type model, even a small displacement of the location of the J-discontinuity can modify the maximum neutral temperature, which in turn has an effect on the fractional abundance profiles: see the r.h.s. upper panel of Figure~\ref{Figure6}.
   
The lower panels of Figure~\ref{Figure6} show the integrated intensities of the total of 156 transitions of para-H$_2$O that result from the three different calculations. In the case of the C-shock, the similarity of the results is striking and extends to very weak lines. The time-scale for the level populations to approach equilibrium is given by $[n(\rm H_2)\it q(J+1 \rightarrow J)]^{\rm{-1}}$ (where $q(J+1 \rightarrow J)$ is the de-excitation rate coefficient for collisions with molecular H$_2$) in the limit of high densities; it is even shorter in the limit of low densities, when radiative decay dominates collisional de-excitation. To validate the assumption of s-e, one must compare this population transfer time-scale with a dynamical time-scale, such as the flow time of the neutral particles through the shock region, that corresponds to the width of the neutral temperature profile shown in the upper panels. Adopting the representative values of 10$^4$~cm$^{-3}$ for the density and 10$^{-10}$~cm$^{-3}$ for the collisional de-excitation rate coefficient results in a population transfer time-scale of 0.03~yr, much less than the flow time of the neutral particles through the C-type shock considered. In the case of the reference CJ-type shock model, this condition is verified for most of the transitions, as the overall agreement between light and dark blue open circles in Figure~\ref{Figure6} confirms. For the brightest transitions, our independent LVG calculation is consistent with that denoted `FP10'. In this case, the discrepancies with the red filled circles are larger, owing to the differences in the water abundance profiles, which are attributable in turn to the temperature differences arising from: the small displacement of J-discontinuity; the treatment of the cooling due to water; and the strength of the grain--gas interaction as the post-shock region is approached. We note that the pattern created by each set of points is similar for the brightest, and hence observable, transitions, and that the pattern is shifted vertically from one set of results to the other. Such a shift is indistinguishable from that arising from the uncertainty in the filling factor, when comparisons are made with observations. 

We conclude from the above analysis that the use of the s-e approximation is justified in the present context of the study of water emission.
   
\subsection{Water emission predictions for BHR71}
\label{sub:wepfBHR71}

As a preamble to this section -- aimed at providing predictions of water line intensities for use in the interpretation of observations -- we study the evolution of the optical depth of the fundamental transitions of both o-H$_2$O and p-H$_2$O in each of our reference models. This evolution, together with the total optical depths, integrated through the shock wave, are shown in Figure~\ref{Figure7}. Owing to the one-dimensional nature of our approach, we probably overestimate the optical depths and find that the transitions are optically thick in most parts of our shock models; this is especially the case of the C-type shock, as the water accumulates in the large post-shock region. The example of the CJ-type shock shows the spread of values of the optical depths in the emitting region -- low when water is dissociated in the high-temperature regime, followed by higher values when water reforms. This figure underlines the requirement for an adequate treatment of the radiative transfer.

As already mentioned, H$_2$, SiO and H$_2$O are believed to have coincident regions of emission, at least to the resolution of currently available observations (e.g. \citealt{Nisini10}). For completeness, we provide predictions of water line intensities for \textit{all} the models that were found to fit the H$_2$ pure rotational excitation diagram acceptably at position of the SiO knot. Given the relative uniformity of the shock conditions in the inner parts of the outflow, revealed by the similarity of the H$_2$ excitation diagrams at the different positions considered, our models should be applicable to water observations at any position. Stationary, C-type shock models are excluded, for the reasons given in Section~\ref{sub:mr}.

The results are given in Tables~\ref{tableb1} and \ref{tableb2} for o-H$_2$O and p-H$_2$O, respectively. The transitions that appear in the Tables are the ones that will be the target of \textit{Herschel} observations. The first three transitions in each table will be observed by HIFI, and the remainder (nine transitions of o-H$_2$O and three of p-H$_2$O) will be observed by PACS, in the framework of the WISH key-program (\citealt{Vandishoeck11}). The calculations comprise the entire set of transitions that are listed on the LAMDA website, and the complete results are available upon request.

The first general comment concerns the dependence of the water emission line intensities on the pre-shock density. We find that the integrated intensities calculated for a pre-shock density of 10$^5$~cm$^{-3}$ are systematically around one order of magnitude larger than those computed in models with pre-shock densities of 10$^4$~cm$^{-3}$, regardless of the values of the other parameters, such as the shock velocity or magnetic-field strength. In addition, and with all other parameters remaining equal, we find that the computed integrated intensities vary by factors of 2--3, and sometimes more, over the range of shock ages considered; this variation may be sufficient, in certain circumstances, to enable the observations to constrain the age of the shock wave.

\begin{figure}
   \centering
   \includegraphics[width=9cm]{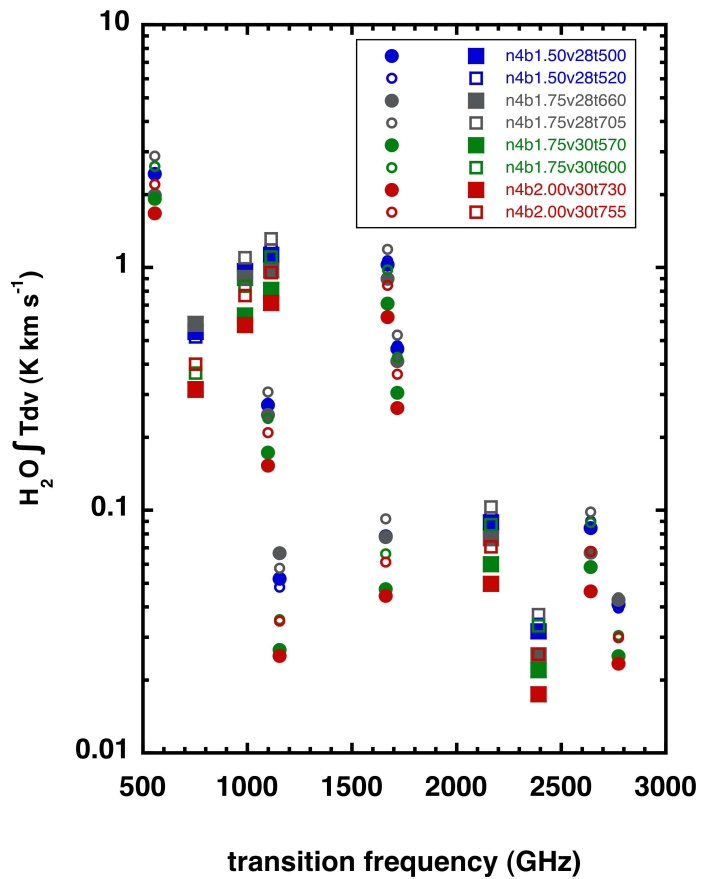}
      \caption{Integrated intensities of the strongest transitions of o- and p-H$_2$O to be targeted by HIFI and PACS, for the eight shock models presented in panels (b) and (c) of Figure~\ref{figure5}. The o-H$_2$O and p-H$_2$O transitions are indicated by circles and squares, respectively. The shock model parameters are abbreviated such that  \lq n4b1.50v28t500', for example, corresponds to a pre-shock density of 10$^4$~cm$^{-3}$, a magnetic-field parameter of 1.5, a shock velocity of 28 km s$^{-1}$, and an age of 500 years.}
         \label{figure8}
   \end{figure}

Finally, we consider the computed data pertaining to the eight shock models that provide the best simultaneous fits for H$_2$ and SiO; these models are shown in Figure~\ref{figure5}, in panels (b) and (c). The results of the calculations of the integrated intensities of the brightest o- and p-H$_2$O transitions, likely to be targetted by HIFI and PACS, are shown in Figure~\ref{figure8}. The filling factor is assumed to be the same and equal to 1 for all the transitions and too small to discriminate between these models. Nevertheless, forthcoming \textit{Herschel} observations should enable us to enhance our understanding of the chemistry of water in interstellar shock waves. 

Water is believed to be the main interstellar reservoir of oxygen, which is the most abundant element in the universe after H and He and plays a major role in interstellar chemistry. Given its rotational energy level structure, water is an important gas coolant and a convenient tracer of shocked regions. Forthcoming observations, through their comparison with the predictions of shock models, should enable us to investigate both the chemical and the physical processes at work in bipolar outflows.

\section{Concluding remarks}
\label{sec:cr}

We have presented a detailed analysis of observations at a few positions in the the bipolar outflow BHR71, assuming that the emission from these regions arises in shock waves. In two cases, namely the Herbig-Haro objects HH320 and HH321, we are able to make use of both the recent \textit{Spitzer} observations of pure rotational transitions and previous, ground-based observations of rovibrational transitions of molecular hydrogen. In order to compare with the predictions of our shock models, we used excitation diagrams. The physical conditions were found to be similar in both cases, with a pre-shock density of 10$^{4}$~cm$^{-3}$ and moderate shock velocities, in the range 15--35~km~s$^{-1}$. The only shock models that fit these data satisfactorily are non-stationary, with ages that depend on the magnetic-field strength and do not exceed 2000~years. Three other positions in the upper and lower lobes of the outflow were studied, using only the pure rotational transitions of H$_2$ observed by \textit{Spitzer}. Our analysis suggests that the excitation conditions at these positions are similar to those in the HH objects and can only be reproduced by non-stationary models.

We present a new map of the SiO(5--4) transition in the inner part of the bipolar outflow BHR71, together with new observations of the SiO(6--5) and SiO(8--7) lines, at one position that is also covered by the \textit{Spitzer} observations and at another position that lies outside of the \textit{Spitzer} maps. At the first position, referred to as \lq the SiO knot', we make use of the combination of data (pure rotational transitions of both H$_2$ and SiO) in order to constrain the parameters of our shock models more tightly than is possible with the rotational transitions of H$_2$ alone. The modelling based on H$_2$ leads to the conclusion as for the other positions in the outflow, namely non-stationary shocks provide the best fits. In particular, shocks with a pre-shock density of 10$^{5}$~cm$^{-3}$, low velocities (10--20~km~s$^{-1}$) and ages (less than 1000~years) were found to be good candidates. Including the SiO emission places constraints on the initial repartition of silicon-bearing material but is of little help in discriminating between the models which fit the H$_2$ rotational excitation diagram. 

We have validated our approach to calculating the water line intensities, which uses the results of shock model as input to a separate radiative transfer code, based on the LVG approximation. To this end, we provide a full comparison between the statistical-equilibrium and the steady-state approximations, based on reference shock models of the stationary, C-type, and non-stationary, CJ-type. We make predictions of water emission line intensities for the SiO knot in the BHR71 outflow. As we expect emission by water to coincide with that by SiO and H$_2$, we provide the results of the LVG calculations of the water line intensities for models that fit satisfactorily both the SiO and H$_2$ emission.\footnote {On request, we can provide water line intensities for all the shock models that fit the H$_2$ observations.} Given that the excitation conditions are relatively uniform in the internal parts of the outflow, we consider that our calculations can be used to compare with observations of water throughout this region.

We shall discuss in a future publication ground-based observations of CO transitions, undertaken with the APEX facility, that could bring further insights into the physical conditions in the BHR71 outflow. Complementary observations of sulphur-bearing molecules and other shock tracers, such as HNCO, are planned. One of the aims of these studies is to investigate the suitability of \lq chemically active outflows', such as BHR71, for observations with the new generation of instruments, such as ALMA.

\begin{acknowledgements}
      BP acknowledges the support of the Deutsche Forschungsgemeinschaft through the Emmy Noether grant PA1692/1--1. We also wish to thank Malcolm Walmsley and an anonymous referee for useful comments that contributed to improving the completeness and clarity of our paper.
\end{acknowledgements}


\bibliographystyle{aa}
\bibliography{biblio}

\newpage

\begin{appendix} 
\section{Model constraints for the HH objects and the SiO knot}

\begin{table}
\caption{Parameters of the models that fit the H$_2$ observations of HH320AB and HH321AB (separated by the double horizontal line). The shock velocities are in km~s$^{-1}$ and the ages in years. Ages between age1 and age2 provide acceptable fits to the observations of H$_2$.}            
\label{tablea1}      
\centering                          
\begin{tabular}{c c c c c c}        
\hline            
\tiny shock type & \tiny log \textit{n$_{\rm H}$} & \tiny b & \tiny velocity & \tiny  age 1 & \tiny age 2 \\    
\hline \hline                       
\tiny CJ & \tiny 4 & \tiny 0.45 & \tiny 15 & \tiny 305 & \tiny 645 \\      
\hline
\tiny CJ & \tiny 4 & \tiny 0.60 & \tiny 15 & \tiny 325 & \tiny 500 \\
\tiny CJ & \tiny 4 & \tiny 0.60 & \tiny 18 & \tiny 215 & \tiny 240 \\
\hline       
\tiny CJ & \tiny 4 & \tiny 0.75 & \tiny 15 & \tiny 470 & \tiny 830 \\
\tiny CJ & \tiny 4 & \tiny 0.75 & \tiny 18 & \tiny 375 & \tiny 375 \\
\tiny CJ & \tiny 4 & \tiny 0.75 & \tiny 25 & \tiny 150 & \tiny 165 \\
\hline           
\tiny CJ & \tiny 4 & \tiny 1.00 & \tiny 15 & \tiny 735 & \tiny 1000 \\
\tiny CJ & \tiny 4 & \tiny 1.00 & \tiny 18 & \tiny 225 & \tiny 690 \\
\hline           
\tiny CJ & \tiny 4 & \tiny 1.25 & \tiny 20 & \tiny 620 & \tiny 780 \\
\tiny CJ & \tiny 4 & \tiny 1.25 & \tiny 28 & \tiny 315 & \tiny 320 \\
\tiny CJ & \tiny 4 & \tiny 1.25 & \tiny 30 & \tiny 260 & \tiny 260 \\
\hline          
\tiny CJ & \tiny 4 & \tiny 1.50 & \tiny 20 & \tiny 900 & \tiny 1040 \\
\tiny CJ & \tiny 4 & \tiny 1.50 & \tiny 28 & \tiny 455 & \tiny 470 \\
\tiny CJ & \tiny 4 & \tiny 1.50 & \tiny 30 & \tiny 380 & \tiny 405 \\
\tiny CJ & \tiny 4 & \tiny 1.50 & \tiny 32 & \tiny 330 & \tiny 340 \\
\tiny CJ & \tiny 4 & \tiny 1.50 & \tiny 35 & \tiny 250 & \tiny 250 \\
\hline          
\tiny CJ & \tiny 4 & \tiny 1.75 & \tiny 20 & \tiny 1125 & \tiny 1295 \\
\tiny CJ & \tiny 4 & \tiny 1.75 & \tiny 22 & \tiny 1015 & \tiny 1045 \\
\tiny CJ & \tiny 4 & \tiny 1.75 & \tiny 28 & \tiny 630 & \tiny 650 \\
\tiny CJ & \tiny 4 & \tiny 1.75 & \tiny 30 & \tiny 540 & \tiny 570 \\
\tiny CJ & \tiny 4 & \tiny 1.75 & \tiny 32 & \tiny 450 & \tiny 465 \\
\tiny CJ & \tiny 4 & \tiny 1.75 & \tiny 35 & \tiny 365 & \tiny 380 \\
\hline
\tiny CJ & \tiny 4 & \tiny 2.00 & \tiny 20 & \tiny 245 & \tiny 1720 \\
\tiny CJ & \tiny 4 & \tiny 2.00 & \tiny 22 & \tiny 675 & \tiny 1390 \\
\tiny CJ & \tiny 4 & \tiny 2.00 & \tiny 28 & \tiny 805 & \tiny 835 \\
\tiny CJ & \tiny 4 & \tiny 2.00 & \tiny 30 & \tiny 695 & \tiny 730 \\
\tiny CJ & \tiny 4 & \tiny 2.00 & \tiny 32 & \tiny 595 & \tiny 650 \\
\tiny CJ & \tiny 4 & \tiny 2.00 & \tiny 35 & \tiny 485 & \tiny 485 \\
\hline \hline                       
\tiny CJ & \tiny 4 & \tiny 0.45 & \tiny 15 & \tiny 220 & \tiny 335 \\    
\tiny CJ & \tiny 4 & \tiny 0.45 & \tiny 18 & \tiny 42 & \tiny 230 \\      
\tiny CJ & \tiny 4 & \tiny 0.45 & \tiny 20 & \tiny 125 & \tiny 175 \\      
\tiny CJ & \tiny 4 & \tiny 0.45 & \tiny 22 & \tiny 81 & \tiny 125 \\      
\hline
\tiny CJ & \tiny 4 & \tiny 0.60 & \tiny 15 & \tiny 325 & \tiny 500 \\
\tiny CJ & \tiny 4 & \tiny 0.60 & \tiny 18 & \tiny 215 & \tiny 325 \\
\tiny CJ & \tiny 4 & \tiny 0.60 & \tiny 20 & \tiny 125 & \tiny 185 \\
\tiny CJ & \tiny 4 & \tiny 0.60 & \tiny 22 & \tiny 130 & \tiny 130 \\
\hline       
\tiny CJ & \tiny 4 & \tiny 0.75 & \tiny 15 & \tiny 470 & \tiny 570 \\
\tiny CJ & \tiny 4 & \tiny 0.75 & \tiny 18 & \tiny 375 & \tiny 375 \\
\tiny CJ & \tiny 4 & \tiny 0.75 & \tiny 20 & \tiny 250 & \tiny 305 \\
\tiny CJ & \tiny 4 & \tiny 0.75 & \tiny 22 & \tiny 225 & \tiny 225 \\
\tiny CJ & \tiny 4 & \tiny 0.75 & \tiny 25 & \tiny 150 & \tiny 165 \\
\hline           
\tiny CJ & \tiny 4 & \tiny 1.00 & \tiny 15 & \tiny 735 & \tiny 915 \\
\tiny CJ & \tiny 4 & \tiny 1.00 & \tiny 18 & \tiny 225 & \tiny 690 \\
\tiny CJ & \tiny 4 & \tiny 1.00 & \tiny 20 & \tiny 430 & \tiny 490 \\
\tiny CJ & \tiny 4 & \tiny 1.00 & \tiny 22 & \tiny 245 & \tiny 350 \\
\tiny CJ & \tiny 4 & \tiny 1.00 & \tiny 25 & \tiny 225 & \tiny 235 \\
\hline           
\tiny CJ & \tiny 4 & \tiny 1.25 & \tiny 20 & \tiny 620 & \tiny 700 \\
\tiny CJ & \tiny 4 & \tiny 1.25 & \tiny 22 & \tiny 525 & \tiny 625 \\
\tiny CJ & \tiny 4 & \tiny 1.25 & \tiny 25 & \tiny 330 & \tiny 425 \\
\hline          
\tiny CJ & \tiny 4 & \tiny 1.50 & \tiny 20 & \tiny 710 & \tiny 900 \\
\tiny CJ & \tiny 4 & \tiny 1.50 & \tiny 22 & \tiny 505 & \tiny 930 \\
\tiny CJ & \tiny 4 & \tiny 1.50 & \tiny 25 & \tiny 355 & \tiny 575 \\
\tiny CJ & \tiny 4 & \tiny 1.50 & \tiny 28 & \tiny 455 & \tiny 470 \\
\tiny CJ & \tiny 4 & \tiny 1.50 & \tiny 30 & \tiny 385 & \tiny 395 \\
\tiny CJ & \tiny 4 & \tiny 1.50 & \tiny 32 & \tiny 330 & \tiny 340 \\
\tiny CJ & \tiny 4 & \tiny 1.50 & \tiny 35 & \tiny 225 & \tiny 260 \\
\hline          
\tiny CJ & \tiny 4 & \tiny 1.75 & \tiny 20 & \tiny 310 & \tiny 1295 \\
\tiny CJ & \tiny 4 & \tiny 1.75 & \tiny 22 & \tiny 315 & \tiny 1045 \\
\tiny CJ & \tiny 4 & \tiny 1.75 & \tiny 25 & \tiny 310 & \tiny 620 \\
\tiny CJ & \tiny 4 & \tiny 1.75 & \tiny 28 & \tiny 650 & \tiny 650 \\
\tiny CJ & \tiny 4 & \tiny 1.75 & \tiny 30 & \tiny 540 & \tiny 540 \\
\tiny CJ & \tiny 4 & \tiny 1.75 & \tiny 32 & \tiny 465 & \tiny 465 \\
\tiny CJ & \tiny 4 & \tiny 1.75 & \tiny 35 & \tiny 355 & \tiny 355 \\
\hline
\tiny CJ & \tiny 4 & \tiny 2.00 & \tiny 20 & \tiny 245 & \tiny 1720 \\
\tiny CJ & \tiny 4 & \tiny 2.00 & \tiny 22 & \tiny 675 & \tiny 1485 \\
\tiny CJ & \tiny 4 & \tiny 2.00 & \tiny 25 & \tiny 565 & \tiny 985 \\
\tiny CJ & \tiny 4 & \tiny 2.00 & \tiny 28 & \tiny 805 & \tiny 835 \\
\tiny CJ & \tiny 4 & \tiny 2.00 & \tiny 30 & \tiny 705 & \tiny 705 \\
\tiny CJ & \tiny 4 & \tiny 2.00 & \tiny 32 & \tiny 605 & \tiny 605 \\
\tiny CJ & \tiny 4 & \tiny 2.00 & \tiny 35 & \tiny 485 & \tiny 485 \\
\hline
\end{tabular}
\end{table}

\begin{table}
\caption{Parameters of the models that fit the H$_2$ observations of the SiO knot. The shock velocities are in km~s$^{-1}$ and the ages in years. Ages between age1 and age2 provide acceptable fits to the observations of H$_2$.}             
\label{tablea2}      
\centering                          
\begin{tabular}{c c c c c c}        
\hline            
shock type & log \textit{n$_{\rm H}$} & b & velocity &  age 1 & age 2 \\    
\hline \hline                       
CJ & 4 & 0.45 & 15 & 335 & 645 \\      
CJ & 4 & 0.45 & 18 & 350 & 425 \\      
CJ & 4 & 0.45 & 20 & 210 & 600 \\      
\hline
CJ & 4 & 0.60 & 15 & 640 & 885 \\
CJ & 4 & 0.60 & 18 & 370 & 730 \\
CJ & 4 & 0.60 & 20 & 260 & 560 \\
CJ & 4 & 0.60 & 22 & 275 & 365 \\
\hline       
CJ & 4 & 0.75 & 15 & 710 & 1120 \\
CJ & 4 & 0.75 & 18 & 375 & 695 \\
CJ & 4 & 0.75 & 20 & 365 & 530 \\
CJ & 4 & 0.75 & 22 & 225 & 450 \\
\hline           
CJ & 4 & 1.00 & 20 & 490 & 740 \\
CJ & 4 & 1.00 & 22 & 370 & 465 \\
CJ & 4 & 1.00 & 25 & 290 & 345 \\
\hline           
CJ & 4 & 1.25 & 22 & 585 & 670 \\
CJ & 4 & 1.25 & 25 & 425 & 530 \\
CJ & 4 & 1.25 & 28 & 325 & 465 \\
\hline          
CJ & 4 & 1.50 & 22 & 830 & 1005 \\
CJ & 4 & 1.50 & 25 & 605 & 755 \\
CJ & 4 & 1.50 & 28 & 500 & 625 \\
CJ & 4 & 1.50 & 30 & 405 & 825 \\
\hline          
CJ & 4 & 1.75 & 22 & 1115 & 1260 \\
CJ & 4 & 1.75 & 25 & 855 & 920 \\
CJ & 4 & 1.75 & 28 & 660 & 890 \\
CJ & 4 & 1.75 & 30 & 570 & 670 \\
\hline
CJ & 4 & 2.00 & 28 & 835 & 920 \\
CJ & 4 & 2.00 & 30 & 730 & 775 \\
\hline                       
CJ & 5 & 0.30 & 10 & 16 & 67 \\      
CJ & 5 & 0.30 & 12 & 23 & 69 \\    
\hline                       
CJ & 5 & 0.45 & 10 & 51 & 115 \\      
CJ & 5 & 0.45 & 12 & 58 & 76 \\        
CJ & 5 & 0.45 & 15 & 24 & 55 \\      
CJ & 5 & 0.45 & 18 & 13 & 35 \\      
\hline
CJ & 5 & 0.60 & 10 & 52 & 155 \\      
CJ & 5 & 0.60 & 12 & 77 & 120 \\        
CJ & 5 & 0.60 & 15 & 53 & 72 \\      
CJ & 5 & 0.60 & 20 & 26 & 62 \\    
\hline
CJ & 5 & 0.75 & 10 & 115 & 250 \\      
CJ & 5 & 0.75 & 12 & 100 & 165 \\        
CJ & 5 & 0.75 & 15 & 86 & 104 \\ 
\hline
CJ & 5 & 1.00 & 10 & 385 & 410 \\      
CJ & 5 & 1.00 & 12 & 205 & 255 \\        
CJ & 5 & 1.00 & 15 & 160 & 175 \\ 
 \hline
CJ & 5 & 1.25 & 15 & 205 & 265 \\          
 \hline
CJ & 5 & 1.50 & 10 & 525 & 775 \\      
CJ & 5 & 1.50 & 15 & 335 & 370 \\                                    
 \hline
CJ & 5 & 1.75 & 10 & 505 & 915 \\      
CJ & 5 & 1.75 & 15 & 315 & 465 \\   
CJ & 5 & 1.75 & 18 & 305 & 340 \\   
 \hline
CJ & 5 & 2.00 & 18 & 435 & 440 \\      
\hline
\end{tabular}
\end{table}

\newpage
\section{Water emission predictions for the SiO knot}

\setcounter{table}{0}
\begin{landscape}
\begin{table}
\begin{center}
\caption{Predicted integrated intensities of the o-H$_2$O lines to be targeted by the HIFI and PACS receivers of the \textit{Herschel} telescope. The units are K km s$^{-1}$. The models considered are those which best fit the H$_2$ pure rotational excitation diagram at the position of  the SiO knot. A model referred to as \lq n4b0.45v15t335' has the following set of input parameters:~a pre-shock density of 10$^4$~cm$^{-3}$, a magnetic-field parameter of 0.45, a shock velocity of 15 km s$^{-1}$, and an age of 335 years.}            
\label{tableb1}      
\begin{tabular}{c c c c c c c c c c c c c }   
\hline                  
\hline
\tiny	transition name	& \tiny	1$_{\rm{10}}$ - 1$_{\rm{01}}$	& \tiny	3$_{\rm{12}}$ - 3$_{\rm{03}}$	& \tiny	3$_{\rm{12}}$ - 2$_{\rm{21}}$	& \tiny	2$_{\rm{21}}$ - 2$_{\rm{12}}$	& \tiny	2$_{\rm{12}}$ - 1$_{\rm{01}}$	& \tiny	3$_{\rm{03}}$ - 2$_{\rm{12}}$	& \tiny	3$_{\rm{30}}$ - 3$_{\rm{21}}$	& \tiny	4$_{\rm{14}}$ - 3$_{\rm{03}}$	& \tiny	2$_{\rm{21}}$ - 1$_{\rm{10}}$	& \tiny	7$_{\rm{07}}$ - 6$_{\rm{16}}$	& \tiny	8$_{\rm{18}}$ - 7$_{\rm{07}}$	& \tiny	9$_{\rm{01}}$ - 8$_{\rm{18}}$	\\
\tiny	frequency (GHz)	& \tiny	556.936	& \tiny	1097.365	& \tiny	1153.127	& \tiny	1661.008	& \tiny	1669.905	& \tiny	1716.770	& \tiny	2196.346	& \tiny	2640.474	& \tiny	2773.977	& \tiny	4166.852	& \tiny	4734.296	& \tiny	5276.520	\\
\hline																										
\hline
\\																										
\tiny	n4b0.45v15t335	& \tiny	5.0E+00	& \tiny	3.6E+00	& \tiny	4.7E-01	& \tiny	6.4E-01	& \tiny	2.1E+00	& \tiny	1.7E+00	& \tiny	1.1E+00	& \tiny	3.9E-01	& \tiny	2.8E-02	& \tiny	9.6E-03	& \tiny	4.0E-03	& \tiny	1.5E-03	\\
\tiny	n4b0.45v15t555	& \tiny	6.6E+00	& \tiny	4.5E+00	& \tiny	5.9E-01	& \tiny	7.8E-01	& \tiny	2.7E+00	& \tiny	2.2E+00	& \tiny	1.4E+00	& \tiny	5.3E-01	& \tiny	3.9E-02	& \tiny	1.5E-02	& \tiny	6.2E-03	& \tiny	2.4E-03	\\
\tiny	n4b0.45v15t645	& \tiny	7.3E+00	& \tiny	4.8E+00	& \tiny	6.3E-01	& \tiny	8.3E-01	& \tiny	2.9E+00	& \tiny	2.3E+00	& \tiny	1.5E+00	& \tiny	5.8E-01	& \tiny	4.2E-02	& \tiny	1.7E-02	& \tiny	7.1E-03	& \tiny	2.7E-03	\\

\tiny	n4b0.45v18t350	& \tiny	7.6E+00	& \tiny	5.8E+00	& \tiny	1.0E+00	& \tiny	1.2E+00	& \tiny	3.8E+00	& \tiny	4.4E+00	& \tiny	2.0E+00	& \tiny	9.7E-01	& \tiny	1.0E-01	& \tiny	2.3E-02	& \tiny	9.2E-03	& \tiny	3.4E-03	\\
\tiny	n4b0.45v18t385	& \tiny	7.9E+00	& \tiny	5.9E+00	& \tiny	1.1E+00	& \tiny	1.3E+00	& \tiny	3.9E+00	& \tiny	4.5E+00	& \tiny	2.1E+00	& \tiny	1.0E+00	& \tiny	1.0E-01	& \tiny	2.4E-02	& \tiny	9.8E-03	& \tiny	3.6E-03	\\
\tiny	n4b0.45v18t425	& \tiny	8.2E+00	& \tiny	6.1E+00	& \tiny	1.1E+00	& \tiny	1.3E+00	& \tiny	4.0E+00	& \tiny	4.6E+00	& \tiny	2.2E+00	& \tiny	1.0E+00	& \tiny	1.1E-01	& \tiny	2.6E-02	& \tiny	1.1E-02	& \tiny	3.9E-03	\\

\tiny	n4b0.45v20t210	& \tiny	7.6E+00	& \tiny	6.3E+00	& \tiny	1.3E+00	& \tiny	1.5E+00	& \tiny	4.4E+00	& \tiny	5.9E+00	& \tiny	2.4E+00	& \tiny	1.3E+00	& \tiny	1.5E-01	& \tiny	2.9E-02	& \tiny	1.1E-02	& \tiny	3.9E-03	\\
\tiny	n4b0.45v20t315	& \tiny	8.8E+00	& \tiny	7.0E+00	& \tiny	1.5E+00	& \tiny	1.6E+00	& \tiny	4.8E+00	& \tiny	6.4E+00	& \tiny	2.7E+00	& \tiny	1.4E+00	& \tiny	1.6E-01	& \tiny	3.5E-02	& \tiny	1.3E-02	& \tiny	4.9E-03	\\
\tiny	n4b0.45v20t600	& \tiny	1.1E+01	& \tiny	7.5E+00	& \tiny	1.4E+00	& \tiny	1.6E+00	& \tiny	5.0E+00	& \tiny	6.0E+00	& \tiny	2.8E+00	& \tiny	1.4E+00	& \tiny	1.5E-01	& \tiny	3.9E-02	& \tiny	1.6E-02	& \tiny	6.0E-03	\\
\\
\hline
\\
\tiny	n4b0.60v15t640	& \tiny	5.9E+00	& \tiny	3.8E+00	& \tiny	4.7E-01	& \tiny	6.5E-01	& \tiny	2.2E+00	& \tiny	1.7E+00	& \tiny	1.1E+00	& \tiny	4.0E-01	& \tiny	2.9E-02	& \tiny	1.1E-02	& \tiny	4.6E-03	& \tiny	1.7E-03	\\
\tiny	n4b0.60v15t655	& \tiny	6.0E+00	& \tiny	3.9E+00	& \tiny	4.8E-01	& \tiny	6.6E-01	& \tiny	2.3E+00	& \tiny	1.8E+00	& \tiny	1.2E+00	& \tiny	4.1E-01	& \tiny	3.0E-02	& \tiny	1.1E-02	& \tiny	4.8E-03	& \tiny	1.8E-03	\\
\tiny	n4b0.60v15t885	& \tiny	7.7E+00	& \tiny	4.6E+00	& \tiny	5.5E-01	& \tiny	7.5E-01	& \tiny	2.6E+00	& \tiny	2.0E+00	& \tiny	1.4E+00	& \tiny	5.1E-01	& \tiny	3.6E-02	& \tiny	1.5E-02	& \tiny	6.4E-03	& \tiny	2.4E-03	\\

\tiny	n4b0.60v18t370	& \tiny	6.6E+00	& \tiny	4.9E+00	& \tiny	8.8E-01	& \tiny	1.1E+00	& \tiny	3.2E+00	& \tiny	3.6E+00	& \tiny	1.7E+00	& \tiny	7.6E-01	& \tiny	8.1E-02	& \tiny	1.7E-02	& \tiny	6.7E-03	& \tiny	2.4E-03	\\
\tiny	n4b0.60v18t515	& \tiny	8.1E+00	& \tiny	5.7E+00	& \tiny	9.7E-01	& \tiny	1.2E+00	& \tiny	3.7E+00	& \tiny	4.0E+00	& \tiny	2.0E+00	& \tiny	8.9E-01	& \tiny	9.1E-02	& \tiny	2.2E-02	& \tiny	9.0E-03	& \tiny	3.4E-03	\\
\tiny	n4b0.60v18t730	& \tiny	1.0E+01	& \tiny	6.7E+00	& \tiny	1.1E+00	& \tiny	1.3E+00	& \tiny	4.2E+00	& \tiny	4.5E+00	& \tiny	2.3E+00	& \tiny	1.1E+00	& \tiny	1.0E-01	& \tiny	3.0E-02	& \tiny	1.2E-02	& \tiny	4.7E-03	\\

\tiny	n4b0.60v20t260	& \tiny	7.1E+00	& \tiny	5.7E+00	& \tiny	1.2E+00	& \tiny	1.4E+00	& \tiny	3.9E+00	& \tiny	5.3E+00	& \tiny	2.2E+00	& \tiny	1.1E+00	& \tiny	1.3E-01	& \tiny	2.3E-02	& \tiny	8.6E-03	& \tiny	3.1E-03	\\
\tiny	n4b0.60v20t330	& \tiny	8.0E+00	& \tiny	6.3E+00	& \tiny	1.3E+00	& \tiny	1.5E+00	& \tiny	4.3E+00	& \tiny	5.7E+00	& \tiny	2.4E+00	& \tiny	1.2E+00	& \tiny	1.4E-01	& \tiny	2.7E-02	& \tiny	1.0E-02	& \tiny	3.8E-03	\\
\tiny	n4b0.60v20t560	& \tiny	1.0E+01	& \tiny	7.3E+00	& \tiny	1.3E+00	& \tiny	1.6E+00	& \tiny	4.8E+00	& \tiny	5.8E+00	& \tiny	2.7E+00	& \tiny	1.3E+00	& \tiny	1.4E-01	& \tiny	3.5E-02	& \tiny	1.4E-02	& \tiny	5.3E-03	\\

\tiny	n4b0.60v22t195	& \tiny	8.5E+00	& \tiny	7.7E+00	& \tiny	1.9E+00	& \tiny	2.0E+00	& \tiny	5.4E+00	& \tiny	8.8E+00	& \tiny	3.2E+00	& \tiny	1.9E+00	& \tiny	2.4E-01	& \tiny	4.5E-02	& \tiny	1.5E-02	& \tiny	5.3E-03	\\
\tiny	n4b0.60v22t275	& \tiny	9.9E+00	& \tiny	8.4E+00	& \tiny	2.0E+00	& \tiny	2.1E+00	& \tiny	5.9E+00	& \tiny	9.4E+00	& \tiny	3.6E+00	& \tiny	2.0E+00	& \tiny	2.6E-01	& \tiny	4.9E-02	& \tiny	1.7E-02	& \tiny	6.1E-03	\\
\tiny	n4b0.60v22t365	& \tiny	1.1E+01	& \tiny	8.6E+00	& \tiny	2.0E+00	& \tiny	2.1E+00	& \tiny	5.9E+00	& \tiny	8.9E+00	& \tiny	3.6E+00	& \tiny	1.9E+00	& \tiny	2.4E-01	& \tiny	4.8E-02	& \tiny	1.8E-02	& \tiny	6.5E-03	\\
\\
\hline
\\
\tiny	n4b0.75v15t710	& \tiny	4.8E+00	& \tiny	3.1E+00	& \tiny	3.5E-01	& \tiny	5.1E-01	& \tiny	1.7E+00	& \tiny	1.3E+00	& \tiny	8.5E-01	& \tiny	2.9E-01	& \tiny	2.0E-02	& \tiny	7.2E-03	& \tiny	3.1E-03	& \tiny	1.1E-03	\\
\tiny	n4b0.75v15t830	& \tiny	5.7E+00	& \tiny	3.5E+00	& \tiny	4.0E-01	& \tiny	5.6E-01	& \tiny	1.9E+00	& \tiny	1.4E+00	& \tiny	9.7E-01	& \tiny	3.3E-01	& \tiny	2.3E-02	& \tiny	9.0E-03	& \tiny	3.8E-03	& \tiny	1.4E-03	\\
\tiny	n4b0.75v15t1120	& \tiny	7.4E+00	& \tiny	4.0E+00	& \tiny	4.2E-01	& \tiny	6.0E-01	& \tiny	2.2E+00	& \tiny	1.5E+00	& \tiny	1.1E+00	& \tiny	3.9E-01	& \tiny	2.4E-02	& \tiny	1.1E-02	& \tiny	5.0E-03	& \tiny	1.9E-03	\\

\tiny	n4b0.75v18t375	& \tiny	4.8E+00	& \tiny	3.7E+00	& \tiny	6.2E-01	& \tiny	7.9E-01	& \tiny	2.3E+00	& \tiny	2.5E+00	& \tiny	1.2E+00	& \tiny	4.9E-01	& \tiny	5.3E-02	& \tiny	9.5E-03	& \tiny	3.7E-03	& \tiny	1.3E-03	\\
\tiny	n4b0.75v18t510	& \tiny	6.4E+00	& \tiny	4.5E+00	& \tiny	7.7E-01	& \tiny	9.6E-01	& \tiny	2.9E+00	& \tiny	3.1E+00	& \tiny	1.5E+00	& \tiny	6.4E-01	& \tiny	6.8E-02	& \tiny	1.4E-02	& \tiny	5.8E-03	& \tiny	2.1E-03	\\
\tiny	n4b0.75v18t695	& \tiny	8.5E+00	& \tiny	5.6E+00	& \tiny	8.8E-01	& \tiny	1.1E+00	& \tiny	3.5E+00	& \tiny	3.6E+00	& \tiny	1.8E+00	& \tiny	8.0E-01	& \tiny	7.8E-02	& \tiny	2.1E-02	& \tiny	8.5E-03	& \tiny	3.2E-03	\\

\tiny	n4b0.75v20t365	& \tiny	7.1E+00	& \tiny	5.5E+00	& \tiny	1.1E+00	& \tiny	1.3E+00	& \tiny	3.7E+00	& \tiny	4.9E+00	& \tiny	2.1E+00	& \tiny	9.9E-01	& \tiny	1.2E-01	& \tiny	2.1E-02	& \tiny	7.8E-03	& \tiny	2.8E-03	\\
\tiny	n4b0.75v20t430	& \tiny	7.9E+00	& \tiny	5.9E+00	& \tiny	1.2E+00	& \tiny	1.4E+00	& \tiny	3.9E+00	& \tiny	5.1E+00	& \tiny	2.2E+00	& \tiny	1.1E+00	& \tiny	1.3E-01	& \tiny	2.4E-02	& \tiny	9.1E-03	& \tiny	3.3E-03	\\
\tiny	n4b0.75v20t530	& \tiny	9.1E+00	& \tiny	6.5E+00	& \tiny	1.2E+00	& \tiny	1.5E+00	& \tiny	4.3E+00	& \tiny	5.4E+00	& \tiny	2.4E+00	& \tiny	1.2E+00	& \tiny	1.3E-01	& \tiny	2.8E-02	& \tiny	1.1E-02	& \tiny	4.1E-03	\\

\tiny	n4b0.75v22t225	& \tiny	7.1E+00	& \tiny	6.4E+00	& \tiny	1.5E+00	& \tiny	1.6E+00	& \tiny	4.5E+00	& \tiny	7.2E+00	& \tiny	2.6E+00	& \tiny	1.5E+00	& \tiny	1.9E-01	& \tiny	3.3E-02	& \tiny	1.1E-02	& \tiny	3.7E-03	\\
\tiny	n4b0.75v22t285	& \tiny	8.4E+00	& \tiny	7.3E+00	& \tiny	1.8E+00	& \tiny	1.9E+00	& \tiny	5.0E+00	& \tiny	8.2E+00	& \tiny	3.1E+00	& \tiny	1.7E+00	& \tiny	2.2E-01	& \tiny	3.8E-02	& \tiny	1.3E-02	& \tiny	4.5E-03	\\
\tiny	n4b0.75v22t450	& \tiny	1.1E+01	& \tiny	8.2E+00	& \tiny	1.8E+00	& \tiny	2.0E+00	& \tiny	5.5E+00	& \tiny	8.1E+00	& \tiny	3.4E+00	& \tiny	1.7E+00	& \tiny	2.2E-01	& \tiny	4.3E-02	& \tiny	1.6E-02	& \tiny	5.8E-03	\\
\\
\hline
\end{tabular}
\end{center}
\end{table}
\end{landscape}

\setcounter{table}{0}
\begin{landscape}
\begin{table}
\begin{center}
\caption{\textit{(continued)} Predicted integrated intensities of the o-H$_2$O lines to be targeted by the HIFI and PACS receivers of the \textit{Herschel} telescope. The units are K km s$^{-1}$. The models considered are those which best fit the H$_2$ pure rotational excitation diagram at the position of  the SiO knot. A model referred to as \lq n4b0.45v15t335' has the following set of input parameters:~a pre-shock density of 10$^4$~cm$^{-3}$, a magnetic-field parameter of 0.45, a shock velocity of 15 km s$^{-1}$, and an age of 335 years.}            
\label{tableb1}      
\begin{tabular}{c c c c c c c c c c c c c }   
\hline                  
\hline
\tiny	transition name	& \tiny	1$_{\rm{10}}$ - 1$_{\rm{01}}$	& \tiny	3$_{\rm{12}}$ - 3$_{\rm{03}}$	& \tiny	3$_{\rm{12}}$ - 2$_{\rm{21}}$	& \tiny	2$_{\rm{21}}$ - 2$_{\rm{12}}$	& \tiny	2$_{\rm{12}}$ - 1$_{\rm{01}}$	& \tiny	3$_{\rm{03}}$ - 2$_{\rm{12}}$	& \tiny	3$_{\rm{30}}$ - 3$_{\rm{21}}$	& \tiny	4$_{\rm{14}}$ - 3$_{\rm{03}}$	& \tiny	2$_{\rm{21}}$ - 1$_{\rm{10}}$	& \tiny	7$_{\rm{07}}$ - 6$_{\rm{16}}$	& \tiny	8$_{\rm{18}}$ - 7$_{\rm{07}}$	& \tiny	9$_{\rm{01}}$ - 8$_{\rm{18}}$	\\
\tiny	frequency (GHz)	& \tiny	556.936	& \tiny	1097.365	& \tiny	1153.127	& \tiny	1661.008	& \tiny	1669.905	& \tiny	1716.770	& \tiny	2196.346	& \tiny	2640.474	& \tiny	2773.977	& \tiny	4166.852	& \tiny	4734.296	& \tiny	5276.520	\\
\hline																										
\hline
\\
\tiny	n4b1.00v20t490	& \tiny	5.3E+00	& \tiny	4.1E+00	& \tiny	8.1E-01	& \tiny	9.7E-01	& \tiny	2.7E+00	& \tiny	3.5E+00	& \tiny	1.5E+00	& \tiny	6.5E-01	& \tiny	8.0E-02	& \tiny	1.2E-02	& \tiny	4.4E-03	& \tiny	1.5E-03	\\
\tiny	n4b1.00v20t605	& \tiny	7.4E+00	& \tiny	5.3E+00	& \tiny	9.9E-01	& \tiny	1.2E+00	& \tiny	3.4E+00	& \tiny	4.2E+00	& \tiny	1.9E+00	& \tiny	8.5E-01	& \tiny	1.0E-01	& \tiny	1.9E-02	& \tiny	7.3E-03	& \tiny	2.6E-03	\\
\tiny	n4b1.00v20t740	& \tiny	8.9E+00	& \tiny	6.0E+00	& \tiny	1.1E+00	& \tiny	1.3E+00	& \tiny	3.8E+00	& \tiny	4.5E+00	& \tiny	2.1E+00	& \tiny	9.7E-01	& \tiny	1.1E-01	& \tiny	2.3E-02	& \tiny	9.3E-03	& \tiny	3.4E-03	\\

\tiny	n4b1.00v22t370	& \tiny	6.0E+00	& \tiny	5.3E+00	& \tiny	1.2E+00	& \tiny	1.3E+00	& \tiny	3.6E+00	& \tiny	5.6E+00	& \tiny	2.1E+00	& \tiny	1.1E+00	& \tiny	1.5E-01	& \tiny	2.1E-02	& \tiny	7.0E-03	& \tiny	2.3E-03	\\
\tiny	n4b1.00v22t435	& \tiny	7.5E+00	& \tiny	6.2E+00	& \tiny	1.5E+00	& \tiny	1.6E+00	& \tiny	4.2E+00	& \tiny	6.6E+00	& \tiny	2.6E+00	& \tiny	1.3E+00	& \tiny	1.8E-01	& \tiny	2.6E-02	& \tiny	9.2E-03	& \tiny	3.2E-03	\\
\tiny	n4b1.00v22t465	& \tiny	8.0E+00	& \tiny	6.4E+00	& \tiny	1.5E+00	& \tiny	1.6E+00	& \tiny	4.3E+00	& \tiny	6.7E+00	& \tiny	2.6E+00	& \tiny	1.3E+00	& \tiny	1.8E-01	& \tiny	2.8E-02	& \tiny	9.8E-03	& \tiny	3.4E-03	\\

\tiny	n4b1.00v25t290	& \tiny	7.3E+00	& \tiny	6.8E+00	& \tiny	1.7E+00	& \tiny	1.8E+00	& \tiny	4.9E+00	& \tiny	8.6E+00	& \tiny	2.9E+00	& \tiny	1.8E+00	& \tiny	2.3E-01	& \tiny	5.2E-02	& \tiny	1.7E-02	& \tiny	5.3E-03	\\
\tiny	n4b1.00v25t320	& \tiny	9.1E+00	& \tiny	8.6E+00	& \tiny	2.4E+00	& \tiny	2.3E+00	& \tiny	6.0E+00	& \tiny	1.2E+01	& \tiny	3.9E+00	& \tiny	2.3E+00	& \tiny	3.3E-01	& \tiny	7.3E-02	& \tiny	2.2E-02	& \tiny	7.1E-03	\\
\tiny	n4b1.00v25t345	& \tiny	1.1E+01	& \tiny	9.9E+00	& \tiny	2.8E+00	& \tiny	2.7E+00	& \tiny	6.8E+00	& \tiny	1.4E+01	& \tiny	4.6E+00	& \tiny	2.7E+00	& \tiny	4.1E-01	& \tiny	8.6E-02	& \tiny	2.6E-02	& \tiny	8.3E-03	\\
\\
\hline
\\
\tiny	n4b1.25v22t585	& \tiny	5.7E+00	& \tiny	4.7E+00	& \tiny	1.0E+00	& \tiny	1.2E+00	& \tiny	3.1E+00	& \tiny	4.6E+00	& \tiny	1.8E+00	& \tiny	8.6E-01	& \tiny	1.2E-01	& \tiny	1.5E-02	& \tiny	5.2E-03	& \tiny	1.7E-03	\\
\tiny	n4b1.25v22t625	& \tiny	6.7E+00	& \tiny	5.2E+00	& \tiny	1.2E+00	& \tiny	1.3E+00	& \tiny	3.5E+00	& \tiny	5.1E+00	& \tiny	2.1E+00	& \tiny	9.7E-01	& \tiny	1.3E-01	& \tiny	1.8E-02	& \tiny	6.5E-03	& \tiny	2.3E-03	\\
\tiny	n4b1.25v22t670	& \tiny	7.3E+00	& \tiny	5.6E+00	& \tiny	1.2E+00	& \tiny	1.4E+00	& \tiny	3.7E+00	& \tiny	5.3E+00	& \tiny	2.2E+00	& \tiny	1.0E+00	& \tiny	1.4E-01	& \tiny	2.0E-02	& \tiny	7.4E-03	& \tiny	2.6E-03	\\

\tiny	n4b1.25v25t425	& \tiny	7.5E+00	& \tiny	7.3E+00	& \tiny	2.0E+00	& \tiny	2.0E+00	& \tiny	5.1E+00	& \tiny	1.0E+01	& \tiny	3.3E+00	& \tiny	1.9E+00	& \tiny	2.8E-01	& \tiny	6.1E-02	& \tiny	1.8E-02	& \tiny	5.6E-03	\\
\tiny	n4b1.25v25t445	& \tiny	8.2E+00	& \tiny	7.9E+00	& \tiny	2.2E+00	& \tiny	2.2E+00	& \tiny	5.4E+00	& \tiny	1.1E+01	& \tiny	3.6E+00	& \tiny	2.1E+00	& \tiny	3.2E-01	& \tiny	6.6E-02	& \tiny	2.0E-02	& \tiny	6.0E-03	\\
\tiny	n4b1.25v25t530	& \tiny	1.1E+01	& \tiny	1.0E+01	& \tiny	2.8E+00	& \tiny	2.8E+00	& \tiny	6.7E+00	& \tiny	1.3E+01	& \tiny	4.7E+00	& \tiny	2.6E+00	& \tiny	4.5E-01	& \tiny	7.6E-02	& \tiny	2.3E-02	& \tiny	7.3E-03	\\

\tiny	n4b1.25v28t325	& \tiny	7.0E-01	& \tiny	2.0E-01	& \tiny	7.4E-03	& \tiny	6.2E-03	& \tiny	9.1E-02	& \tiny	6.9E-03	& \tiny	4.8E-02	& \tiny	2.2E-02	& \tiny	6.0E-05	& \tiny	1.3E-03	& \tiny	6.4E-04	& \tiny	2.7E-04	\\
\tiny	n4b1.25v28t465	& \tiny	4.8E+00	& \tiny	2.0E+00	& \tiny	5.9E-02	& \tiny	1.5E-01	& \tiny	9.3E-01	& \tiny	7.9E-02	& \tiny	5.3E-01	& \tiny	2.0E-01	& \tiny	8.1E-04	& \tiny	1.2E-02	& \tiny	5.9E-03	& \tiny	2.5E-03	\\
\\
\hline
\\
\tiny	n4b1.50v22t830	& \tiny	4.9E+00	& \tiny	4.0E+00	& \tiny	8.2E-01	& \tiny	9.6E-01	& \tiny	2.6E+00	& \tiny	3.6E+00	& \tiny	1.4E+00	& \tiny	6.3E-01	& \tiny	8.5E-02	& \tiny	1.0E-02	& \tiny	3.7E-03	& \tiny	1.3E-03	\\
\tiny	n4b1.50v22t930	& \tiny	6.9E+00	& \tiny	5.0E+00	& \tiny	1.0E+00	& \tiny	1.2E+00	& \tiny	3.2E+00	& \tiny	4.3E+00	& \tiny	1.9E+00	& \tiny	8.2E-01	& \tiny	1.1E-01	& \tiny	1.6E-02	& \tiny	6.0E-03	& \tiny	2.1E-03	\\
\tiny	n4b1.50v22t1005	& \tiny	8.2E+00	& \tiny	5.6E+00	& \tiny	1.1E+00	& \tiny	1.3E+00	& \tiny	3.5E+00	& \tiny	4.6E+00	& \tiny	2.1E+00	& \tiny	9.0E-01	& \tiny	1.1E-01	& \tiny	2.0E-02	& \tiny	7.5E-03	& \tiny	2.7E-03	\\

\tiny	n4b1.50v25t605	& \tiny	7.1E+00	& \tiny	6.9E+00	& \tiny	1.9E+00	& \tiny	1.9E+00	& \tiny	4.7E+00	& \tiny	9.3E+00	& \tiny	3.1E+00	& \tiny	1.7E+00	& \tiny	2.6E-01	& \tiny	4.6E-02	& \tiny	1.4E-02	& \tiny	4.1E-03	\\
\tiny	n4b1.50v25t655	& \tiny	8.4E+00	& \tiny	7.7E+00	& \tiny	2.1E+00	& \tiny	2.1E+00	& \tiny	5.2E+00	& \tiny	1.0E+01	& \tiny	3.5E+00	& \tiny	1.9E+00	& \tiny	3.1E-01	& \tiny	5.0E-02	& \tiny	1.5E-02	& \tiny	4.6E-03	\\
\tiny	n4b1.50v25t755	& \tiny	1.1E+01	& \tiny	8.9E+00	& \tiny	2.3E+00	& \tiny	2.4E+00	& \tiny	5.9E+00	& \tiny	1.1E+01	& \tiny	4.0E+00	& \tiny	2.1E+00	& \tiny	3.5E-01	& \tiny	5.1E-02	& \tiny	1.7E-02	& \tiny	5.5E-03	\\

\tiny	n4b1.50v28t500	& \tiny	2.4E+00	& \tiny	1.0E+00	& \tiny	4.1E-02	& \tiny	7.8E-02	& \tiny	4.6E-01	& \tiny	5.2E-02	& \tiny	2.7E-01	& \tiny	8.4E-02	& \tiny	4.7E-04	& \tiny	3.9E-03	& \tiny	1.9E-03	& \tiny	7.5E-04	\\
\tiny	n4b1.50v28t520	& \tiny	2.6E+00	& \tiny	1.1E+00	& \tiny	4.0E-02	& \tiny	7.9E-02	& \tiny	4.8E-01	& \tiny	4.8E-02	& \tiny	2.8E-01	& \tiny	9.0E-02	& \tiny	4.5E-04	& \tiny	4.5E-03	& \tiny	2.1E-03	& \tiny	8.6E-04	\\
\tiny	n4b1.50v28t625	& \tiny	4.7E+00	& \tiny	2.0E+00	& \tiny	5.6E-02	& \tiny	1.5E-01	& \tiny	8.9E-01	& \tiny	7.7E-02	& \tiny	5.2E-01	& \tiny	1.9E-01	& \tiny	7.7E-04	& \tiny	1.1E-02	& \tiny	5.3E-03	& \tiny	2.2E-03	\\

\tiny	n4b1.50v30t405	& \tiny	1.8E+00	& \tiny	6.6E-01	& \tiny	2.4E-02	& \tiny	4.2E-02	& \tiny	2.9E-01	& \tiny	2.5E-02	& \tiny	1.6E-01	& \tiny	5.7E-02	& \tiny	2.6E-04	& \tiny	2.8E-03	& \tiny	1.4E-03	& \tiny	5.6E-04	\\
\tiny	n4b1.50v30t435	& \tiny	2.2E+00	& \tiny	8.0E-01	& \tiny	2.6E-02	& \tiny	5.0E-02	& \tiny	3.5E-01	& \tiny	2.9E-02	& \tiny	2.0E-01	& \tiny	7.5E-02	& \tiny	2.9E-04	& \tiny	4.2E-03	& \tiny	2.1E-03	& \tiny	8.7E-04	\\
\tiny	n4b1.50v30t825	& \tiny	1.3E+01	& \tiny	6.0E+00	& \tiny	1.6E-01	& \tiny	4.8E-01	& \tiny	2.8E+00	& \tiny	2.5E-01	& \tiny	1.7E+00	& \tiny	6.3E-01	& \tiny	2.6E-03	& \tiny	3.9E-02	& \tiny	1.9E-02	& \tiny	8.0E-03	\\
\\
\hline
\end{tabular}
\end{center}
\end{table}
\end{landscape}

\setcounter{table}{0}
\begin{landscape}
\begin{table}
\begin{center}
\caption{\textit{(continued)} Predicted integrated intensities of the o-H$_2$O lines to be targeted by the HIFI and PACS receivers of the \textit{Herschel} telescope. The units are K km s$^{-1}$. The models considered are those which best fit the H$_2$ pure rotational excitation diagram at the position of  the SiO knot. A model referred to as \lq n4b0.45v15t335' has the following set of input parameters:~a pre-shock density of 10$^4$~cm$^{-3}$, a magnetic-field parameter of 0.45, a shock velocity of 15 km s$^{-1}$, and an age of 335 years.}            
\label{tableb1}      
\begin{tabular}{c c c c c c c c c c c c c }   
\hline                  
\hline
\tiny	transition name	& \tiny	1$_{\rm{10}}$ - 1$_{\rm{01}}$	& \tiny	3$_{\rm{12}}$ - 3$_{\rm{03}}$	& \tiny	3$_{\rm{12}}$ - 2$_{\rm{21}}$	& \tiny	2$_{\rm{21}}$ - 2$_{\rm{12}}$	& \tiny	2$_{\rm{12}}$ - 1$_{\rm{01}}$	& \tiny	3$_{\rm{03}}$ - 2$_{\rm{12}}$	& \tiny	3$_{\rm{30}}$ - 3$_{\rm{21}}$	& \tiny	4$_{\rm{14}}$ - 3$_{\rm{03}}$	& \tiny	2$_{\rm{21}}$ - 1$_{\rm{10}}$	& \tiny	7$_{\rm{07}}$ - 6$_{\rm{16}}$	& \tiny	8$_{\rm{18}}$ - 7$_{\rm{07}}$	& \tiny	9$_{\rm{01}}$ - 8$_{\rm{18}}$	\\
\tiny	frequency (GHz)	& \tiny	556.936	& \tiny	1097.365	& \tiny	1153.127	& \tiny	1661.008	& \tiny	1669.905	& \tiny	1716.770	& \tiny	2196.346	& \tiny	2640.474	& \tiny	2773.977	& \tiny	4166.852	& \tiny	4734.296	& \tiny	5276.520	\\
\hline																										
\hline
\\
\tiny	n4b1.75v22t1115	& \tiny	4.5E+00	& \tiny	3.5E+00	& \tiny	7.0E-01	& \tiny	8.3E-01	& \tiny	2.2E+00	& \tiny	3.0E+00	& \tiny	1.2E+00	& \tiny	5.2E-01	& \tiny	7.0E-02	& \tiny	8.5E-03	& \tiny	3.0E-03	& \tiny	1.0E-03	\\
\tiny	n4b1.75v22t1180	& \tiny	5.5E+00	& \tiny	4.0E+00	& \tiny	7.6E-01	& \tiny	9.3E-01	& \tiny	2.5E+00	& \tiny	3.2E+00	& \tiny	1.4E+00	& \tiny	5.8E-01	& \tiny	7.5E-02	& \tiny	1.1E-02	& \tiny	4.0E-03	& \tiny	1.4E-03	\\
\tiny	n4b1.75v22t1260	& \tiny	6.8E+00	& \tiny	4.6E+00	& \tiny	8.5E-01	& \tiny	1.0E+00	& \tiny	2.9E+00	& \tiny	3.6E+00	& \tiny	1.6E+00	& \tiny	6.8E-01	& \tiny	8.4E-02	& \tiny	1.4E-02	& \tiny	5.3E-03	& \tiny	1.8E-03	\\

\tiny	n4b1.75v25t855	& \tiny	7.3E+00	& \tiny	6.5E+00	& \tiny	1.7E+00	& \tiny	1.8E+00	& \tiny	4.4E+00	& \tiny	8.3E+00	& \tiny	2.9E+00	& \tiny	1.5E+00	& \tiny	2.4E-01	& \tiny	3.3E-02	& \tiny	1.0E-02	& \tiny	3.2E-03	\\
\tiny	n4b1.75v25t885	& \tiny	7.9E+00	& \tiny	6.8E+00	& \tiny	1.8E+00	& \tiny	1.9E+00	& \tiny	4.6E+00	& \tiny	8.4E+00	& \tiny	3.0E+00	& \tiny	1.6E+00	& \tiny	2.5E-01	& \tiny	3.4E-02	& \tiny	1.0E-02	& \tiny	3.4E-03	\\
\tiny	n4b1.75v25t920	& \tiny	8.7E+00	& \tiny	7.2E+00	& \tiny	1.8E+00	& \tiny	1.9E+00	& \tiny	4.8E+00	& \tiny	8.5E+00	& \tiny	3.2E+00	& \tiny	1.6E+00	& \tiny	2.6E-01	& \tiny	3.5E-02	& \tiny	1.1E-02	& \tiny	3.7E-03	\\

\tiny	n4b1.75v28t660	& \tiny	2.0E+00	& \tiny	9.0E-01	& \tiny	4.3E-02	& \tiny	7.8E-02	& \tiny	4.1E-01	& \tiny	6.7E-02	& \tiny	2.5E-01	& \tiny	6.7E-02	& \tiny	5.4E-04	& \tiny	2.5E-03	& \tiny	1.1E-03	& \tiny	4.3E-04	\\
\tiny	n4b1.75v28t705	& \tiny	2.9E+00	& \tiny	1.2E+00	& \tiny	4.4E-02	& \tiny	9.2E-02	& \tiny	5.3E-01	& \tiny	5.8E-02	& \tiny	3.1E-01	& \tiny	9.8E-02	& \tiny	5.1E-04	& \tiny	4.7E-03	& \tiny	2.2E-03	& \tiny	8.7E-04	\\
\tiny	n4b1.75v28t890	& \tiny	1.5E+01	& \tiny	1.2E+01	& \tiny	2.7E+00	& \tiny	2.9E+00	& \tiny	7.4E+00	& \tiny	1.3E+01	& \tiny	5.1E+00	& \tiny	2.7E+00	& \tiny	4.2E-01	& \tiny	7.7E-02	& \tiny	2.5E-02	& \tiny	8.5E-03	\\

\tiny	n4b1.75v30t570	& \tiny	1.9E+00	& \tiny	7.1E-01	& \tiny	2.5E-02	& \tiny	4.7E-02	& \tiny	3.0E-01	& \tiny	2.7E-02	& \tiny	1.7E-01	& \tiny	5.8E-02	& \tiny	2.6E-04	& \tiny	2.9E-03	& \tiny	1.4E-03	& \tiny	5.5E-04	\\
\tiny	n4b1.75v30t600	& \tiny	2.6E+00	& \tiny	9.8E-01	& \tiny	3.0E-02	& \tiny	6.6E-02	& \tiny	4.3E-01	& \tiny	3.5E-02	& \tiny	2.4E-01	& \tiny	8.9E-02	& \tiny	3.5E-04	& \tiny	4.9E-03	& \tiny	2.4E-03	& \tiny	9.7E-04	\\
\tiny	n4b1.75v30t670	& \tiny	4.3E+00	& \tiny	1.8E+00	& \tiny	5.0E-02	& \tiny	1.3E-01	& \tiny	8.1E-01	& \tiny	7.0E-02	& \tiny	4.7E-01	& \tiny	1.8E-01	& \tiny	7.0E-04	& \tiny	1.0E-02	& \tiny	5.0E-03	& \tiny	2.0E-03	\\
\\
\hline
\\
\tiny	n4b2.00v28t835	& \tiny	1.8E+00	& \tiny	1.1E+00	& \tiny	8.3E-02	& \tiny	1.3E-01	& \tiny	5.5E-01	& \tiny	2.3E-01	& \tiny	3.1E-01	& \tiny	8.3E-02	& \tiny	2.3E-03	& \tiny	2.1E-03	& \tiny	8.4E-04	& \tiny	2.8E-04	\\
\tiny	n4b2.00v28t880	& \tiny	2.8E+00	& \tiny	1.5E+00	& \tiny	1.1E-01	& \tiny	1.8E-01	& \tiny	7.7E-01	& \tiny	3.1E-01	& \tiny	4.2E-01	& \tiny	1.3E-01	& \tiny	3.4E-03	& \tiny	4.0E-03	& \tiny	1.7E-03	& \tiny	6.3E-04	\\
\tiny	n4b2.00v28t920	& \tiny	5.2E+00	& \tiny	3.3E+00	& \tiny	3.9E-01	& \tiny	5.5E-01	& \tiny	1.9E+00	& \tiny	1.5E+00	& \tiny	9.5E-01	& \tiny	3.8E-01	& \tiny	3.2E-02	& \tiny	1.0E-02	& \tiny	4.2E-03	& \tiny	1.5E-03	\\

\tiny	n4b2.00v30t730	& \tiny	1.7E+00	& \tiny	6.3E-01	& \tiny	2.3E-02	& \tiny	4.4E-02	& \tiny	2.6E-01	& \tiny	2.5E-02	& \tiny	1.5E-01	& \tiny	4.6E-02	& \tiny	2.4E-04	& \tiny	2.0E-03	& \tiny	9.2E-04	& \tiny	3.5E-04	\\
\tiny	n4b2.00v30t755	& \tiny	2.2E+00	& \tiny	8.4E-01	& \tiny	3.0E-02	& \tiny	6.1E-02	& \tiny	3.6E-01	& \tiny	3.5E-02	& \tiny	2.1E-01	& \tiny	6.8E-02	& \tiny	3.2E-04	& \tiny	3.2E-03	& \tiny	1.5E-03	& \tiny	5.9E-04	\\
\tiny	n4b2.00v30t775	& \tiny	2.6E+00	& \tiny	9.9E-01	& \tiny	3.1E-02	& \tiny	7.0E-02	& \tiny	4.3E-01	& \tiny	3.7E-02	& \tiny	2.4E-01	& \tiny	8.6E-02	& \tiny	3.5E-04	& \tiny	4.4E-03	& \tiny	2.1E-03	& \tiny	8.5E-04	\\
\\
\hline
\\
\tiny	n5b0.30v10t16	& \tiny	1.3E+01	& \tiny	1.4E+01	& \tiny	2.2E+00	& \tiny	2.3E+00	& \tiny	9.9E+00	& \tiny	7.7E+00	& \tiny	6.1E+00	& \tiny	2.6E+00	& \tiny	1.5E-01	& \tiny	8.2E-02	& \tiny	3.3E-02	& \tiny	1.3E-02	\\
\tiny	n5b0.30v10t33	& \tiny	1.5E+01	& \tiny	1.5E+01	& \tiny	2.7E+00	& \tiny	2.9E+00	& \tiny	1.1E+01	& \tiny	1.0E+01	& \tiny	6.9E+00	& \tiny	3.1E+00	& \tiny	2.2E-01	& \tiny	8.6E-02	& \tiny	3.4E-02	& \tiny	1.3E-02	\\
\tiny	n5b0.30v10t67	& \tiny	1.8E+01	& \tiny	1.6E+01	& \tiny	2.9E+00	& \tiny	3.2E+00	& \tiny	1.2E+01	& \tiny	1.1E+01	& \tiny	7.3E+00	& \tiny	3.3E+00	& \tiny	2.4E-01	& \tiny	9.0E-02	& \tiny	3.6E-02	& \tiny	1.4E-02	\\

\tiny	n5b0.30v12t23	& \tiny	1.6E+01	& \tiny	1.7E+01	& \tiny	3.7E+00	& \tiny	3.8E+00	& \tiny	1.3E+01	& \tiny	1.6E+01	& \tiny	8.0E+00	& \tiny	4.6E+00	& \tiny	4.3E-01	& \tiny	1.3E-01	& \tiny	4.7E-02	& \tiny	1.7E-02	\\
\tiny	n5b0.30v12t48	& \tiny	2.2E+01	& \tiny	2.1E+01	& \tiny	4.6E+00	& \tiny	4.6E+00	& \tiny	1.6E+01	& \tiny	1.9E+01	& \tiny	9.8E+00	& \tiny	5.8E+00	& \tiny	5.3E-01	& \tiny	1.7E-01	& \tiny	6.6E-02	& \tiny	2.5E-02	\\
\tiny	n5b0.30v12t69	& \tiny	2.4E+01	& \tiny	2.1E+01	& \tiny	4.4E+00	& \tiny	4.4E+00	& \tiny	1.6E+01	& \tiny	1.8E+01	& \tiny	9.9E+00	& \tiny	5.6E+00	& \tiny	4.8E-01	& \tiny	1.8E-01	& \tiny	6.9E-02	& \tiny	2.7E-02	\\
\\
\hline
\\
\tiny	n5b0.45v10t51	& \tiny	1.4E+01	& \tiny	1.4E+01	& \tiny	2.5E+00	& \tiny	2.7E+00	& \tiny	9.9E+00	& \tiny	9.4E+00	& \tiny	6.0E+00	& \tiny	2.7E+00	& \tiny	2.0E-01	& \tiny	7.1E-02	& \tiny	2.8E-02	& \tiny	1.1E-02	\\
\tiny	n5b0.45v10t88	& \tiny	1.5E+01	& \tiny	1.4E+01	& \tiny	2.6E+00	& \tiny	2.9E+00	& \tiny	1.0E+01	& \tiny	9.9E+00	& \tiny	6.2E+00	& \tiny	2.8E+00	& \tiny	2.2E-01	& \tiny	7.2E-02	& \tiny	2.9E-02	& \tiny	1.1E-02	\\
\tiny	n5b0.45v10t115	& \tiny	1.6E+01	& \tiny	1.5E+01	& \tiny	2.6E+00	& \tiny	2.9E+00	& \tiny	1.0E+01	& \tiny	1.0E+01	& \tiny	6.3E+00	& \tiny	2.8E+00	& \tiny	2.2E-01	& \tiny	7.4E-02	& \tiny	3.0E-02	& \tiny	1.1E-02	\\

\tiny	n5b0.45v12t58	& \tiny	1.5E+01	& \tiny	1.5E+01	& \tiny	3.4E+00	& \tiny	3.6E+00	& \tiny	1.2E+01	& \tiny	1.4E+01	& \tiny	7.0E+00	& \tiny	3.9E+00	& \tiny	3.7E-01	& \tiny	1.0E-01	& \tiny	3.8E-02	& \tiny	1.4E-02	\\
\tiny	n5b0.45v12t76	& \tiny	1.8E+01	& \tiny	1.7E+01	& \tiny	3.7E+00	& \tiny	3.8E+00	& \tiny	1.3E+01	& \tiny	1.5E+01	& \tiny	7.7E+00	& \tiny	4.3E+00	& \tiny	4.0E-01	& \tiny	1.2E-01	& \tiny	4.6E-02	& \tiny	1.7E-02	\\

\tiny	n5b0.45v15t24	& \tiny	1.4E+01	& \tiny	1.5E+01	& \tiny	3.6E+00	& \tiny	3.4E+00	& \tiny	1.1E+01	& \tiny	1.7E+01	& \tiny	7.1E+00	& \tiny	4.7E+00	& \tiny	5.0E-01	& \tiny	1.5E-01	& \tiny	5.0E-02	& \tiny	1.8E-02	\\
\tiny	n5b0.45v15t40	& \tiny	2.5E+01	& \tiny	2.6E+01	& \tiny	7.3E+00	& \tiny	6.8E+00	& \tiny	2.1E+01	& \tiny	3.6E+01	& \tiny	1.4E+01	& \tiny	9.5E+00	& \tiny	1.2E+00	& \tiny	3.9E-01	& \tiny	1.3E-01	& \tiny	4.6E-02	\\
\tiny	n5b0.45v15t55	& \tiny	3.3E+01	& \tiny	3.2E+01	& \tiny	8.3E+00	& \tiny	7.8E+00	& \tiny	2.5E+01	& \tiny	4.0E+01	& \tiny	1.6E+01	& \tiny	1.1E+01	& \tiny	1.4E+00	& \tiny	5.0E-01	& \tiny	1.7E-01	& \tiny	6.5E-02	\\

\tiny	n5b0.45v18t13	& \tiny	1.6E+01	& \tiny	1.8E+01	& \tiny	4.7E+00	& \tiny	4.3E+00	& \tiny	1.5E+01	& \tiny	2.4E+01	& \tiny	9.6E+00	& \tiny	7.4E+00	& \tiny	7.7E-01	& \tiny	4.4E-01	& \tiny	1.5E-01	& \tiny	5.4E-02	\\
\tiny	n5b0.45v18t24	& \tiny	3.5E+01	& \tiny	4.2E+01	& \tiny	1.4E+01	& \tiny	1.2E+01	& \tiny	3.4E+01	& \tiny	7.6E+01	& \tiny	2.2E+01	& \tiny	1.9E+01	& \tiny	2.9E+00	& \tiny	2.1E+00	& \tiny	6.4E-01	& \tiny	2.0E-01	\\
\tiny	n5b0.45v18t35	& \tiny	4.8E+01	& \tiny	5.4E+01	& \tiny	1.7E+01	& \tiny	1.5E+01	& \tiny	4.3E+01	& \tiny	8.9E+01	& \tiny	2.9E+01	& \tiny	2.3E+01	& \tiny	4.1E+00	& \tiny	2.4E+00	& \tiny	6.8E-01	& \tiny	2.1E-01	\\
\\
\hline
\end{tabular}
\end{center}
\end{table}
\end{landscape}

\setcounter{table}{0}
\begin{landscape}
\begin{table}
\begin{center}
\caption{\textit{(continued)} Predicted integrated intensities of the o-H$_2$O lines to be targeted by the HIFI and PACS receivers of the \textit{Herschel} telescope. The units are K km s$^{-1}$. The models considered are those which best fit the H$_2$ pure rotational excitation diagram at the position of  the SiO knot. A model referred to as \lq n4b0.45v15t335' has the following set of input parameters:~a pre-shock density of 10$^4$~cm$^{-3}$, a magnetic-field parameter of 0.45, a shock velocity of 15 km s$^{-1}$, and an age of 335 years.}            
\label{tableb1}      
\begin{tabular}{c c c c c c c c c c c c c }   
\hline                  
\hline
\tiny	transition name	& \tiny	1$_{\rm{10}}$ - 1$_{\rm{01}}$	& \tiny	3$_{\rm{12}}$ - 3$_{\rm{03}}$	& \tiny	3$_{\rm{12}}$ - 2$_{\rm{21}}$	& \tiny	2$_{\rm{21}}$ - 2$_{\rm{12}}$	& \tiny	2$_{\rm{12}}$ - 1$_{\rm{01}}$	& \tiny	3$_{\rm{03}}$ - 2$_{\rm{12}}$	& \tiny	3$_{\rm{30}}$ - 3$_{\rm{21}}$	& \tiny	4$_{\rm{14}}$ - 3$_{\rm{03}}$	& \tiny	2$_{\rm{21}}$ - 1$_{\rm{10}}$	& \tiny	7$_{\rm{07}}$ - 6$_{\rm{16}}$	& \tiny	8$_{\rm{18}}$ - 7$_{\rm{07}}$	& \tiny	9$_{\rm{01}}$ - 8$_{\rm{18}}$	\\
\tiny	frequency (GHz)	& \tiny	556.936	& \tiny	1097.365	& \tiny	1153.127	& \tiny	1661.008	& \tiny	1669.905	& \tiny	1716.770	& \tiny	2196.346	& \tiny	2640.474	& \tiny	2773.977	& \tiny	4166.852	& \tiny	4734.296	& \tiny	5276.520	\\
\hline																										
\hline
\\
\tiny	n5b0.60v10t52	& \tiny	1.3E+01	& \tiny	1.2E+01	& \tiny	2.1E+00	& \tiny	2.3E+00	& \tiny	8.7E+00	& \tiny	7.8E+00	& \tiny	5.0E+00	& \tiny	2.2E+00	& \tiny	1.6E-01	& \tiny	6.0E-02	& \tiny	2.4E-02	& \tiny	9.2E-03	\\
\tiny	n5b0.60v10t105	& \tiny	1.4E+01	& \tiny	1.3E+01	& \tiny	2.3E+00	& \tiny	2.6E+00	& \tiny	9.2E+00	& \tiny	8.9E+00	& \tiny	5.4E+00	& \tiny	2.4E+00	& \tiny	1.9E-01	& \tiny	6.1E-02	& \tiny	2.5E-02	& \tiny	9.3E-03	\\
\tiny	n5b0.60v10t155	& \tiny	1.4E+01	& \tiny	1.3E+01	& \tiny	2.4E+00	& \tiny	2.7E+00	& \tiny	9.3E+00	& \tiny	9.1E+00	& \tiny	5.5E+00	& \tiny	2.4E+00	& \tiny	1.9E-01	& \tiny	6.2E-02	& \tiny	2.5E-02	& \tiny	9.4E-03	\\

\tiny	n5b0.60v12t77	& \tiny	1.3E+01	& \tiny	1.3E+01	& \tiny	2.6E+00	& \tiny	2.8E+00	& \tiny	9.4E+00	& \tiny	1.1E+01	& \tiny	5.6E+00	& \tiny	2.8E+00	& \tiny	2.5E-01	& \tiny	7.1E-02	& \tiny	2.7E-02	& \tiny	1.0E-02	\\
\tiny	n5b0.60v12t85	& \tiny	1.3E+01	& \tiny	1.3E+01	& \tiny	2.6E+00	& \tiny	2.8E+00	& \tiny	9.4E+00	& \tiny	1.1E+01	& \tiny	5.6E+00	& \tiny	2.8E+00	& \tiny	2.6E-01	& \tiny	7.1E-02	& \tiny	2.8E-02	& \tiny	1.0E-02	\\
\tiny	n5b0.60v12t120	& \tiny	1.7E+01	& \tiny	1.5E+01	& \tiny	3.2E+00	& \tiny	3.4E+00	& \tiny	1.1E+01	& \tiny	1.3E+01	& \tiny	6.7E+00	& \tiny	3.6E+00	& \tiny	3.4E-01	& \tiny	9.7E-02	& \tiny	3.7E-02	& \tiny	1.4E-02	\\

\tiny	n5b0.60v15t53	& \tiny	1.5E+01	& \tiny	1.5E+01	& \tiny	3.9E+00	& \tiny	3.8E+00	& \tiny	1.1E+01	& \tiny	1.9E+01	& \tiny	7.4E+00	& \tiny	4.8E+00	& \tiny	5.5E-01	& \tiny	1.4E-01	& \tiny	4.5E-02	& \tiny	1.6E-02	\\
\tiny	n5b0.60v15t72	& \tiny	2.5E+01	& \tiny	2.5E+01	& \tiny	6.7E+00	& \tiny	6.4E+00	& \tiny	1.9E+01	& \tiny	3.2E+01	& \tiny	1.3E+01	& \tiny	8.4E+00	& \tiny	1.1E+00	& \tiny	3.2E-01	& \tiny	1.1E-01	& \tiny	3.9E-02	\\

\tiny	n5b0.60v20t26	& \tiny	1.1E+01	& \tiny	1.3E+01	& \tiny	4.0E+00	& \tiny	3.4E+00	& \tiny	1.0E+01	& \tiny	2.3E+01	& \tiny	6.6E+00	& \tiny	5.9E+00	& \tiny	7.5E-01	& \tiny	5.9E-01	& \tiny	2.0E-01	& \tiny	6.8E-02	\\
\tiny	n5b0.60v20t34	& \tiny	3.1E+01	& \tiny	3.5E+01	& \tiny	1.2E+01	& \tiny	1.0E+01	& \tiny	2.7E+01	& \tiny	6.8E+01	& \tiny	1.8E+01	& \tiny	1.5E+01	& \tiny	3.3E+00	& \tiny	3.0E+00	& \tiny	1.1E+00	& \tiny	3.1E-01	\\
\tiny	n5b0.60v20t62	& \tiny	6.2E+01	& \tiny	7.0E+01	& \tiny	2.3E+01	& \tiny	2.0E+01	& \tiny	5.5E+01	& \tiny	1.2E+02	& \tiny	3.7E+01	& \tiny	2.9E+01	& \tiny	6.9E+00	& \tiny	4.8E+00	& \tiny	1.5E+00	& \tiny	4.0E-01	\\
\\
\hline
\\
\tiny	n5b0.75v10t115	& \tiny	1.3E+01	& \tiny	1.2E+01	& \tiny	2.1E+00	& \tiny	2.4E+00	& \tiny	8.4E+00	& \tiny	7.9E+00	& \tiny	4.8E+00	& \tiny	2.1E+00	& \tiny	1.6E-01	& \tiny	5.4E-02	& \tiny	2.2E-02	& \tiny	8.2E-03	\\
\tiny	n5b0.75v10t220	& \tiny	1.3E+01	& \tiny	1.2E+01	& \tiny	2.2E+00	& \tiny	2.5E+00	& \tiny	8.5E+00	& \tiny	8.4E+00	& \tiny	5.0E+00	& \tiny	2.2E+00	& \tiny	1.8E-01	& \tiny	5.4E-02	& \tiny	2.2E-02	& \tiny	8.2E-03	\\
\tiny	n5b0.75v10t250	& \tiny	1.3E+01	& \tiny	1.2E+01	& \tiny	2.2E+00	& \tiny	2.5E+00	& \tiny	8.5E+00	& \tiny	8.4E+00	& \tiny	5.0E+00	& \tiny	2.2E+00	& \tiny	1.8E-01	& \tiny	5.4E-02	& \tiny	2.2E-02	& \tiny	8.2E-03	\\

\tiny	n5b0.75v12t100	& \tiny	1.3E+01	& \tiny	1.2E+01	& \tiny	2.4E+00	& \tiny	2.7E+00	& \tiny	8.9E+00	& \tiny	9.8E+00	& \tiny	5.2E+00	& \tiny	2.6E+00	& \tiny	2.3E-01	& \tiny	6.6E-02	& \tiny	2.6E-02	& \tiny	9.6E-03	\\
\tiny	n5b0.75v12t140	& \tiny	1.3E+01	& \tiny	1.2E+01	& \tiny	2.5E+00	& \tiny	2.7E+00	& \tiny	9.0E+00	& \tiny	1.0E+01	& \tiny	5.3E+00	& \tiny	2.6E+00	& \tiny	2.4E-01	& \tiny	6.6E-02	& \tiny	2.6E-02	& \tiny	9.6E-03	\\
\tiny	n5b0.75v12t165	& \tiny	1.4E+01	& \tiny	1.3E+01	& \tiny	2.7E+00	& \tiny	3.0E+00	& \tiny	9.6E+00	& \tiny	1.1E+01	& \tiny	5.7E+00	& \tiny	2.9E+00	& \tiny	2.7E-01	& \tiny	7.4E-02	& \tiny	2.9E-02	& \tiny	1.1E-02	\\

\tiny	n5b0.75v15t86	& \tiny	1.4E+01	& \tiny	1.4E+01	& \tiny	3.5E+00	& \tiny	3.5E+00	& \tiny	1.0E+01	& \tiny	1.6E+01	& \tiny	6.5E+00	& \tiny	4.0E+00	& \tiny	4.8E-01	& \tiny	1.0E-01	& \tiny	3.5E-02	& \tiny	1.2E-02	\\
\tiny	n5b0.75v15t104	& \tiny	2.1E+01	& \tiny	2.1E+01	& \tiny	5.7E+00	& \tiny	5.5E+00	& \tiny	1.6E+01	& \tiny	2.7E+01	& \tiny	1.1E+01	& \tiny	6.8E+00	& \tiny	9.0E-01	& \tiny	2.3E-01	& \tiny	7.7E-02	& \tiny	2.8E-02	\\
\\
\hline
\\
\tiny	n5b1.00v10t385	& \tiny	1.3E+01	& \tiny	1.1E+01	& \tiny	1.9E+00	& \tiny	2.3E+00	& \tiny	7.6E+00	& \tiny	7.5E+00	& \tiny	4.4E+00	& \tiny	1.8E+00	& \tiny	1.5E-01	& \tiny	4.5E-02	& \tiny	1.8E-02	& \tiny	6.7E-03	\\
\tiny	n5b1.00v10t410	& \tiny	1.3E+01	& \tiny	1.1E+01	& \tiny	1.9E+00	& \tiny	2.3E+00	& \tiny	7.7E+00	& \tiny	7.5E+00	& \tiny	4.4E+00	& \tiny	1.8E+00	& \tiny	1.5E-01	& \tiny	4.4E-02	& \tiny	1.8E-02	& \tiny	6.6E-03	\\

\tiny	n5b1.00v12t205	& \tiny	1.3E+01	& \tiny	1.2E+01	& \tiny	2.3E+00	& \tiny	2.6E+00	& \tiny	8.5E+00	& \tiny	9.3E+00	& \tiny	4.9E+00	& \tiny	2.4E+00	& \tiny	2.1E-01	& \tiny	6.0E-02	& \tiny	2.4E-02	& \tiny	9.0E-03	\\
\tiny	n5b1.00v12t235	& \tiny	1.3E+01	& \tiny	1.2E+01	& \tiny	2.3E+00	& \tiny	2.6E+00	& \tiny	8.4E+00	& \tiny	9.3E+00	& \tiny	4.9E+00	& \tiny	2.4E+00	& \tiny	2.1E-01	& \tiny	6.0E-02	& \tiny	2.4E-02	& \tiny	8.9E-03	\\
\tiny	n5b1.00v12t255	& \tiny	1.3E+01	& \tiny	1.2E+01	& \tiny	2.3E+00	& \tiny	2.6E+00	& \tiny	8.4E+00	& \tiny	9.3E+00	& \tiny	4.9E+00	& \tiny	2.4E+00	& \tiny	2.1E-01	& \tiny	6.0E-02	& \tiny	2.4E-02	& \tiny	8.9E-03	\\

\tiny	n5b1.00v15t160	& \tiny	1.3E+01	& \tiny	1.2E+01	& \tiny	3.1E+00	& \tiny	3.2E+00	& \tiny	9.2E+00	& \tiny	1.4E+01	& \tiny	5.8E+00	& \tiny	3.4E+00	& \tiny	4.0E-01	& \tiny	8.0E-02	& \tiny	2.8E-02	& \tiny	1.0E-02	\\
\tiny	n5b1.00v15t175	& \tiny	1.6E+01	& \tiny	1.5E+01	& \tiny	3.9E+00	& \tiny	4.0E+00	& \tiny	1.1E+01	& \tiny	1.8E+01	& \tiny	7.3E+00	& \tiny	4.4E+00	& \tiny	5.4E-01	& \tiny	1.2E-01	& \tiny	4.0E-02	& \tiny	1.4E-02	\\
\\
\hline
\\
\tiny	n5b1.25v15t205	& \tiny	1.2E+01	& \tiny	1.1E+01	& \tiny	2.6E+00	& \tiny	2.8E+00	& \tiny	8.1E+00	& \tiny	1.2E+01	& \tiny	4.9E+00	& \tiny	2.7E+00	& \tiny	3.1E-01	& \tiny	6.2E-02	& \tiny	2.2E-02	& \tiny	8.1E-03	\\
\tiny	n5b1.25v15t265	& \tiny	1.4E+01	& \tiny	1.3E+01	& \tiny	3.3E+00	& \tiny	3.5E+00	& \tiny	9.7E+00	& \tiny	1.5E+01	& \tiny	6.0E+00	& \tiny	3.5E+00	& \tiny	4.2E-01	& \tiny	8.7E-02	& \tiny	3.1E-02	& \tiny	1.1E-02	\\
\\
\hline
\\
\tiny	n5b1.50v10t525	& \tiny	1.1E+01	& \tiny	9.8E+00	& \tiny	1.6E+00	& \tiny	2.0E+00	& \tiny	6.4E+00	& \tiny	6.0E+00	& \tiny	3.5E+00	& \tiny	1.4E+00	& \tiny	1.1E-01	& \tiny	3.1E-02	& \tiny	1.2E-02	& \tiny	4.4E-03	\\
\tiny	n5b1.50v10t775	& \tiny	1.1E+01	& \tiny	9.7E+00	& \tiny	1.6E+00	& \tiny	2.0E+00	& \tiny	6.4E+00	& \tiny	6.1E+00	& \tiny	3.5E+00	& \tiny	1.4E+00	& \tiny	1.2E-01	& \tiny	3.0E-02	& \tiny	1.2E-02	& \tiny	4.4E-03	\\

\tiny	n5b1.50v15t335	& \tiny	1.1E+01	& \tiny	1.1E+01	& \tiny	2.4E+00	& \tiny	2.7E+00	& \tiny	7.7E+00	& \tiny	1.1E+01	& \tiny	4.6E+00	& \tiny	2.5E+00	& \tiny	2.7E-01	& \tiny	5.6E-02	& \tiny	2.1E-02	& \tiny	7.5E-03	\\
\tiny	n5b1.50v15t370	& \tiny	1.2E+01	& \tiny	1.1E+01	& \tiny	2.6E+00	& \tiny	2.8E+00	& \tiny	8.0E+00	& \tiny	1.1E+01	& \tiny	4.8E+00	& \tiny	2.6E+00	& \tiny	3.0E-01	& \tiny	6.0E-02	& \tiny	2.2E-02	& \tiny	8.0E-03	\\
\\
\hline
\end{tabular}
\end{center}
\end{table}
\end{landscape}

\setcounter{table}{0}
\begin{landscape}
\begin{table}
\begin{center}
\caption{\textit{(continued)} Predicted integrated intensities of the o-H$_2$O lines to be targeted by the HIFI and PACS receivers of the \textit{Herschel} telescope. The units are K km s$^{-1}$. The models considered are those which best fit the H$_2$ pure rotational excitation diagram at the position of  the SiO knot. A model referred to as \lq n4b0.45v15t335' has the following set of input parameters:~a pre-shock density of 10$^4$~cm$^{-3}$, a magnetic-field parameter of 0.45, a shock velocity of 15 km s$^{-1}$, and an age of 335 years.}            
\label{tableb1}      
\begin{tabular}{c c c c c c c c c c c c c }   
\hline                  
\hline
\tiny	transition name	& \tiny	1$_{\rm{10}}$ - 1$_{\rm{01}}$	& \tiny	3$_{\rm{12}}$ - 3$_{\rm{03}}$	& \tiny	3$_{\rm{12}}$ - 2$_{\rm{21}}$	& \tiny	2$_{\rm{21}}$ - 2$_{\rm{12}}$	& \tiny	2$_{\rm{12}}$ - 1$_{\rm{01}}$	& \tiny	3$_{\rm{03}}$ - 2$_{\rm{12}}$	& \tiny	3$_{\rm{30}}$ - 3$_{\rm{21}}$	& \tiny	4$_{\rm{14}}$ - 3$_{\rm{03}}$	& \tiny	2$_{\rm{21}}$ - 1$_{\rm{10}}$	& \tiny	7$_{\rm{07}}$ - 6$_{\rm{16}}$	& \tiny	8$_{\rm{18}}$ - 7$_{\rm{07}}$	& \tiny	9$_{\rm{01}}$ - 8$_{\rm{18}}$	\\
\tiny	frequency (GHz)	& \tiny	556.936	& \tiny	1097.365	& \tiny	1153.127	& \tiny	1661.008	& \tiny	1669.905	& \tiny	1716.770	& \tiny	2196.346	& \tiny	2640.474	& \tiny	2773.977	& \tiny	4166.852	& \tiny	4734.296	& \tiny	5276.520	\\
\hline																										
\hline
\\
\tiny	n5b1.75v10t505	& \tiny	1.0E+01	& \tiny	9.1E+00	& \tiny	1.4E+00	& \tiny	1.8E+00	& \tiny	5.8E+00	& \tiny	5.3E+00	& \tiny	3.1E+00	& \tiny	1.2E+00	& \tiny	9.8E-02	& \tiny	2.5E-02	& \tiny	9.9E-03	& \tiny	3.5E-03	\\
\tiny	n5b1.75v10t915	& \tiny	1.0E+01	& \tiny	9.1E+00	& \tiny	1.4E+00	& \tiny	1.9E+00	& \tiny	5.8E+00	& \tiny	5.5E+00	& \tiny	3.2E+00	& \tiny	1.2E+00	& \tiny	1.0E-01	& \tiny	2.5E-02	& \tiny	9.7E-03	& \tiny	3.4E-03	\\

\tiny	n5b1.75v12t315	& \tiny	1.1E+01	& \tiny	1.0E+01	& \tiny	2.3E+00	& \tiny	2.5E+00	& \tiny	7.5E+00	& \tiny	9.7E+00	& \tiny	4.3E+00	& \tiny	2.3E+00	& \tiny	2.4E-01	& \tiny	5.3E-02	& \tiny	2.0E-02	& \tiny	7.4E-03	\\
\tiny	n5b1.75v12t410	& \tiny	1.1E+01	& \tiny	1.0E+01	& \tiny	2.3E+00	& \tiny	2.5E+00	& \tiny	7.5E+00	& \tiny	9.9E+00	& \tiny	4.3E+00	& \tiny	2.3E+00	& \tiny	2.5E-01	& \tiny	5.3E-02	& \tiny	2.0E-02	& \tiny	7.3E-03	\\
\tiny	n5b1.75v12t465	& \tiny	1.1E+01	& \tiny	1.0E+01	& \tiny	2.3E+00	& \tiny	2.6E+00	& \tiny	7.5E+00	& \tiny	9.9E+00	& \tiny	4.4E+00	& \tiny	2.3E+00	& \tiny	2.5E-01	& \tiny	5.3E-02	& \tiny	2.0E-02	& \tiny	7.3E-03	\\

\tiny	n5b1.75v18t305	& \tiny	1.3E+01	& \tiny	1.2E+01	& \tiny	3.4E+00	& \tiny	3.4E+00	& \tiny	8.8E+00	& \tiny	1.6E+01	& \tiny	6.0E+00	& \tiny	3.5E+00	& \tiny	4.8E-01	& \tiny	8.6E-02	& \tiny	2.7E-02	& \tiny	9.3E-03	\\
\tiny	n5b1.75v18t325	& \tiny	1.3E+01	& \tiny	1.2E+01	& \tiny	3.4E+00	& \tiny	3.4E+00	& \tiny	8.9E+00	& \tiny	1.6E+01	& \tiny	6.1E+00	& \tiny	3.5E+00	& \tiny	4.9E-01	& \tiny	8.7E-02	& \tiny	2.8E-02	& \tiny	9.4E-03	\\
\tiny	n5b1.75v18t340	& \tiny	1.4E+01	& \tiny	1.3E+01	& \tiny	3.8E+00	& \tiny	3.7E+00	& \tiny	9.6E+00	& \tiny	1.8E+01	& \tiny	6.6E+00	& \tiny	3.9E+00	& \tiny	5.7E-01	& \tiny	1.0E-01	& \tiny	3.2E-02	& \tiny	1.1E-02	\\
\\
\hline
\\
\tiny	n5b2.00v18t435	& \tiny	1.2E+01	& \tiny	1.1E+01	& \tiny	3.0E+00	& \tiny	3.1E+00	& \tiny	8.1E+00	& \tiny	1.4E+01	& \tiny	5.4E+00	& \tiny	3.0E+00	& \tiny	4.0E-01	& \tiny	7.0E-02	& \tiny	2.3E-02	& \tiny	8.0E-03	\\
\tiny	n5b2.00v18t440	& \tiny	1.2E+01	& \tiny	1.1E+01	& \tiny	3.1E+00	& \tiny	3.2E+00	& \tiny	8.3E+00	& \tiny	1.4E+01	& \tiny	5.5E+00	& \tiny	3.1E+00	& \tiny	4.2E-01	& \tiny	7.3E-02	& \tiny	2.4E-02	& \tiny	8.4E-03	\\
\\
\hline    
\end{tabular}
\end{center}
\end{table}
\end{landscape}

\setcounter{table}{1}
\begin{table*}
\begin{center}
\caption{Predicted integrated intensities of the p-H$_2$O lines to be targeted by the HIFI and PACS receivers of the \textit{Herschel} telescope. The units are K km s$^{-1}$. The models considered are those which best fit the H$_2$ pure rotational excitation diagram at the position of  the SiO knot. A model referred to as \lq n4b0.45v15t335' has the following set of input parameters:~a pre-shock density of 10$^4$~cm$^{-3}$, a magnetic-field parameter of 0.45, a shock velocity of 15 km s$^{-1}$, and an age of 335 years.}            
\label{tableb2}      
\begin{tabular}{c c c c c c c c c c c c c }   
\hline                  
\hline
\tiny	transition name	& \tiny	1$_{\rm{11}}$ - 0$_{\rm{00}}$	& \tiny	2$_{\rm{02}}$ - 1$_{\rm{11}}$	& \tiny	2$_{\rm{11}}$ - 2$_{\rm{02}}$	& \tiny	3$_{\rm{13}}$ - 2$_{\rm{02}}$	& \tiny	4$_{\rm{04}}$ - 3$_{\rm{13}}$	& \tiny	3$_{\rm{22}}$ - 2$_{\rm{11}}$	\\
\tiny	frequency (GHz)	& \tiny	1113.343	& \tiny	987.927	& \tiny	752.033	& \tiny	2164.132	& \tiny	2391.573	& \tiny	3331.458	\\
\hline														
\hline
\\														
\tiny	n4b0.45v15t335	& \tiny	3.3E+00	& \tiny	4.2E+00	& \tiny	3.4E+00	& \tiny	5.4E-01	& \tiny	1.9E-02	& \tiny	1.4E-01	\\
\tiny	n4b0.45v15t555	& \tiny	4.1E+00	& \tiny	5.1E+00	& \tiny	3.9E+00	& \tiny	7.1E-01	& \tiny	2.6E-02	& \tiny	2.0E-01	\\
\tiny	n4b0.45v15t645	& \tiny	4.5E+00	& \tiny	5.5E+00	& \tiny	4.2E+00	& \tiny	7.6E-01	& \tiny	2.8E-02	& \tiny	2.1E-01	\\

\tiny	n4b0.45v18t350	& \tiny	4.9E+00	& \tiny	6.6E+00	& \tiny	5.1E+00	& \tiny	1.2E+00	& \tiny	6.9E-02	& \tiny	3.7E-01	\\
\tiny	n4b0.45v18t385	& \tiny	5.1E+00	& \tiny	6.7E+00	& \tiny	5.2E+00	& \tiny	1.2E+00	& \tiny	7.1E-02	& \tiny	3.8E-01	\\
\tiny	n4b0.45v18t425	& \tiny	5.2E+00	& \tiny	6.9E+00	& \tiny	5.3E+00	& \tiny	1.3E+00	& \tiny	7.5E-02	& \tiny	4.0E-01	\\

\tiny	n4b0.45v20t210	& \tiny	5.2E+00	& \tiny	7.2E+00	& \tiny	5.7E+00	& \tiny	1.5E+00	& \tiny	1.2E-01	& \tiny	5.1E-01	\\
\tiny	n4b0.45v20t315	& \tiny	5.8E+00	& \tiny	7.8E+00	& \tiny	6.1E+00	& \tiny	1.7E+00	& \tiny	1.3E-01	& \tiny	5.8E-01	\\
\tiny	n4b0.45v20t600	& \tiny	6.5E+00	& \tiny	8.2E+00	& \tiny	6.1E+00	& \tiny	1.6E+00	& \tiny	1.2E-01	& \tiny	5.7E-01	\\
\\
\hline
\\
\tiny	n4b0.60v15t640	& \tiny	3.6E+00	& \tiny	4.4E+00	& \tiny	3.4E+00	& \tiny	5.6E-01	& \tiny	1.8E-02	& \tiny	1.5E-01	\\
\tiny	n4b0.60v15t655	& \tiny	3.6E+00	& \tiny	4.5E+00	& \tiny	3.4E+00	& \tiny	5.7E-01	& \tiny	1.9E-02	& \tiny	1.5E-01	\\
\tiny	n4b0.60v15t885	& \tiny	4.4E+00	& \tiny	5.1E+00	& \tiny	3.8E+00	& \tiny	6.7E-01	& \tiny	2.2E-02	& \tiny	1.9E-01	\\

\tiny	n4b0.60v18t370	& \tiny	4.3E+00	& \tiny	5.7E+00	& \tiny	4.5E+00	& \tiny	9.8E-01	& \tiny	5.2E-02	& \tiny	2.9E-01	\\
\tiny	n4b0.60v18t515	& \tiny	5.0E+00	& \tiny	6.4E+00	& \tiny	4.9E+00	& \tiny	1.1E+00	& \tiny	6.0E-02	& \tiny	3.4E-01	\\
\tiny	n4b0.60v18t730	& \tiny	6.0E+00	& \tiny	7.2E+00	& \tiny	5.3E+00	& \tiny	1.3E+00	& \tiny	7.3E-02	& \tiny	4.1E-01	\\

\tiny	n4b0.60v20t260	& \tiny	4.8E+00	& \tiny	6.5E+00	& \tiny	5.2E+00	& \tiny	1.3E+00	& \tiny	1.0E-01	& \tiny	4.5E-01	\\
\tiny	n4b0.60v20t330	& \tiny	5.3E+00	& \tiny	6.9E+00	& \tiny	5.6E+00	& \tiny	1.4E+00	& \tiny	1.1E-01	& \tiny	4.9E-01	\\
\tiny	n4b0.60v20t560	& \tiny	6.4E+00	& \tiny	7.9E+00	& \tiny	6.0E+00	& \tiny	1.5E+00	& \tiny	1.1E-01	& \tiny	5.4E-01	\\

\tiny	n4b0.60v22t195	& \tiny	6.0E+00	& \tiny	8.4E+00	& \tiny	7.1E+00	& \tiny	2.0E+00	& \tiny	2.3E-01	& \tiny	8.3E-01	\\
\tiny	n4b0.60v22t275	& \tiny	6.8E+00	& \tiny	9.0E+00	& \tiny	7.6E+00	& \tiny	2.2E+00	& \tiny	2.5E-01	& \tiny	9.0E-01	\\
\tiny	n4b0.60v22t365	& \tiny	7.0E+00	& \tiny	9.1E+00	& \tiny	7.5E+00	& \tiny	2.1E+00	& \tiny	2.3E-01	& \tiny	8.7E-01	\\
\\
\hline
\\
\tiny	n4b0.75v15t710	& \tiny	2.9E+00	& \tiny	3.6E+00	& \tiny	2.8E+00	& \tiny	4.1E-01	& \tiny	1.2E-02	& \tiny	1.1E-01	\\
\tiny	n4b0.75v15t830	& \tiny	3.3E+00	& \tiny	4.0E+00	& \tiny	3.1E+00	& \tiny	4.7E-01	& \tiny	1.4E-02	& \tiny	1.2E-01	\\
\tiny	n4b0.75v15t1120	& \tiny	3.9E+00	& \tiny	4.4E+00	& \tiny	3.2E+00	& \tiny	5.2E-01	& \tiny	1.5E-02	& \tiny	1.4E-01	\\

\tiny	n4b0.75v18t375	& \tiny	3.2E+00	& \tiny	4.3E+00	& \tiny	3.5E+00	& \tiny	6.8E-01	& \tiny	3.1E-02	& \tiny	1.8E-01	\\
\tiny	n4b0.75v18t510	& \tiny	4.0E+00	& \tiny	5.2E+00	& \tiny	4.1E+00	& \tiny	8.5E-01	& \tiny	4.1E-02	& \tiny	2.4E-01	\\
\tiny	n4b0.75v18t695	& \tiny	5.0E+00	& \tiny	6.2E+00	& \tiny	4.7E+00	& \tiny	1.0E+00	& \tiny	5.0E-02	& \tiny	3.1E-01	\\

\tiny	n4b0.75v20t365	& \tiny	4.7E+00	& \tiny	6.1E+00	& \tiny	5.0E+00	& \tiny	1.2E+00	& \tiny	8.9E-02	& \tiny	4.0E-01	\\
\tiny	n4b0.75v20t430	& \tiny	5.0E+00	& \tiny	6.5E+00	& \tiny	5.2E+00	& \tiny	1.3E+00	& \tiny	9.5E-02	& \tiny	4.3E-01	\\
\tiny	n4b0.75v20t530	& \tiny	5.6E+00	& \tiny	7.0E+00	& \tiny	5.5E+00	& \tiny	1.4E+00	& \tiny	1.0E-01	& \tiny	4.7E-01	\\

\tiny	n4b0.75v22t225	& \tiny	5.0E+00	& \tiny	7.0E+00	& \tiny	6.0E+00	& \tiny	1.6E+00	& \tiny	1.8E-01	& \tiny	6.6E-01	\\
\tiny	n4b0.75v22t285	& \tiny	5.8E+00	& \tiny	7.8E+00	& \tiny	6.7E+00	& \tiny	1.9E+00	& \tiny	2.1E-01	& \tiny	7.6E-01	\\
\tiny	n4b0.75v22t450	& \tiny	6.9E+00	& \tiny	8.6E+00	& \tiny	7.1E+00	& \tiny	2.0E+00	& \tiny	2.1E-01	& \tiny	7.8E-01	\\
\\
\hline
\\
\tiny	n4b1.00v20t490	& \tiny	3.5E+00	& \tiny	4.7E+00	& \tiny	4.0E+00	& \tiny	8.5E-01	& \tiny	5.2E-02	& \tiny	2.6E-01	\\
\tiny	n4b1.00v20t605	& \tiny	4.6E+00	& \tiny	5.8E+00	& \tiny	4.7E+00	& \tiny	1.1E+00	& \tiny	6.9E-02	& \tiny	3.4E-01	\\
\tiny	n4b1.00v20t740	& \tiny	5.3E+00	& \tiny	6.4E+00	& \tiny	5.0E+00	& \tiny	1.2E+00	& \tiny	7.8E-02	& \tiny	3.9E-01	\\

\tiny	n4b1.00v22t370	& \tiny	4.2E+00	& \tiny	5.8E+00	& \tiny	5.0E+00	& \tiny	1.3E+00	& \tiny	1.2E-01	& \tiny	4.8E-01	\\
\tiny	n4b1.00v22t435	& \tiny	5.1E+00	& \tiny	6.6E+00	& \tiny	5.8E+00	& \tiny	1.5E+00	& \tiny	1.6E-01	& \tiny	5.8E-01	\\
\tiny	n4b1.00v22t465	& \tiny	5.3E+00	& \tiny	6.8E+00	& \tiny	5.9E+00	& \tiny	1.5E+00	& \tiny	1.6E-01	& \tiny	6.0E-01	\\

\tiny	n4b1.00v25t290	& \tiny	5.2E+00	& \tiny	7.5E+00	& \tiny	6.4E+00	& \tiny	1.9E+00	& \tiny	2.3E-01	& \tiny	8.3E-01	\\
\tiny	n4b1.00v25t320	& \tiny	6.5E+00	& \tiny	8.8E+00	& \tiny	8.2E+00	& \tiny	2.5E+00	& \tiny	3.6E-01	& \tiny	1.2E+00	\\
\tiny	n4b1.00v25t345	& \tiny	7.5E+00	& \tiny	9.8E+00	& \tiny	9.5E+00	& \tiny	2.8E+00	& \tiny	4.6E-01	& \tiny	1.5E+00	\\
\\
\hline
\\
\tiny	n4b1.25v22t585	& \tiny	3.9E+00	& \tiny	5.1E+00	& \tiny	4.5E+00	& \tiny	1.0E+00	& \tiny	8.9E-02	& \tiny	3.7E-01	\\
\tiny	n4b1.25v22t625	& \tiny	4.4E+00	& \tiny	5.7E+00	& \tiny	4.9E+00	& \tiny	1.2E+00	& \tiny	1.0E-01	& \tiny	4.2E-01	\\
\tiny	n4b1.25v22t670	& \tiny	4.7E+00	& \tiny	5.9E+00	& \tiny	5.1E+00	& \tiny	1.2E+00	& \tiny	1.1E-01	& \tiny	4.5E-01	\\

\tiny	n4b1.25v25t425	& \tiny	5.5E+00	& \tiny	7.4E+00	& \tiny	7.2E+00	& \tiny	2.0E+00	& \tiny	3.1E-01	& \tiny	1.0E+00	\\
\tiny	n4b1.25v25t445	& \tiny	6.0E+00	& \tiny	7.8E+00	& \tiny	7.7E+00	& \tiny	2.2E+00	& \tiny	3.4E-01	& \tiny	1.1E+00	\\
\tiny	n4b1.25v25t530	& \tiny	7.8E+00	& \tiny	9.5E+00	& \tiny	9.6E+00	& \tiny	2.7E+00	& \tiny	4.6E-01	& \tiny	1.5E+00	\\

\tiny	n4b1.25v28t325	& \tiny	2.3E-01	& \tiny	1.6E-01	& \tiny	5.3E-02	& \tiny	2.0E-02	& \tiny	2.4E-04	& \tiny	8.1E-03	\\
\tiny	n4b1.25v28t465	& \tiny	2.2E+00	& \tiny	1.7E+00	& \tiny	7.5E-01	& \tiny	1.9E-01	& \tiny	3.2E-03	& \tiny	7.5E-02	\\
\\
\hline
\end{tabular}
\end{center}
\end{table*}

\setcounter{table}{1}
\begin{table*}
\begin{center}
\caption{\textit{(continued)} Predicted integrated intensities of the p-H$_2$O lines to be targeted by the HIFI and PACS receivers of the \textit{Herschel} telescope. The units are K km s$^{-1}$. The models considered are those which best fit the H$_2$ pure rotational excitation diagram at the position of  the SiO knot. A model referred to as \lq n4b0.45v15t335' has the following set of input parameters:~a pre-shock density of 10$^4$~cm$^{-3}$, a magnetic-field parameter of 0.45, a shock velocity of 15 km s$^{-1}$, and an age of 335 years.}            
\label{tableb2}      
\begin{tabular}{c c c c c c c c c c c c c }   
\hline                  
\hline
\tiny	transition name	& \tiny	1$_{\rm{11}}$ - 0$_{\rm{00}}$	& \tiny	2$_{\rm{02}}$ - 1$_{\rm{11}}$	& \tiny	2$_{\rm{11}}$ - 2$_{\rm{02}}$	& \tiny	3$_{\rm{13}}$ - 2$_{\rm{02}}$	& \tiny	4$_{\rm{04}}$ - 3$_{\rm{13}}$	& \tiny	3$_{\rm{22}}$ - 2$_{\rm{11}}$	\\
\tiny	frequency (GHz)	& \tiny	1113.343	& \tiny	987.927	& \tiny	752.033	& \tiny	2164.132	& \tiny	2391.573	& \tiny	3331.458	\\
\hline														
\hline
\\	
\tiny	n4b1.50v22t830	& \tiny	3.3E+00	& \tiny	4.4E+00	& \tiny	3.9E+00	& \tiny	8.2E-01	& \tiny	5.8E-02	& \tiny	2.6E-01	\\
\tiny	n4b1.50v22t930	& \tiny	4.3E+00	& \tiny	5.4E+00	& \tiny	4.6E+00	& \tiny	1.0E+00	& \tiny	7.7E-02	& \tiny	3.5E-01	\\
\tiny	n4b1.50v22t1005	& \tiny	5.0E+00	& \tiny	5.9E+00	& \tiny	4.9E+00	& \tiny	1.1E+00	& \tiny	8.1E-02	& \tiny	3.8E-01	\\

\tiny	n4b1.50v25t605	& \tiny	5.3E+00	& \tiny	6.9E+00	& \tiny	6.9E+00	& \tiny	1.8E+00	& \tiny	2.7E-01	& \tiny	9.1E-01	\\
\tiny	n4b1.50v25t655	& \tiny	6.0E+00	& \tiny	7.5E+00	& \tiny	7.6E+00	& \tiny	2.0E+00	& \tiny	3.1E-01	& \tiny	1.0E+00	\\
\tiny	n4b1.50v25t755	& \tiny	7.3E+00	& \tiny	8.6E+00	& \tiny	8.3E+00	& \tiny	2.2E+00	& \tiny	3.4E-01	& \tiny	1.1E+00	\\

\tiny	n4b1.50v28t500	& \tiny	1.1E+00	& \tiny	9.7E-01	& \tiny	5.4E-01	& \tiny	8.9E-02	& \tiny	2.0E-03	& \tiny	3.2E-02	\\
\tiny	n4b1.50v28t520	& \tiny	1.2E+00	& \tiny	9.8E-01	& \tiny	5.2E-01	& \tiny	9.3E-02	& \tiny	1.9E-03	& \tiny	3.4E-02	\\
\tiny	n4b1.50v28t625	& \tiny	2.2E+00	& \tiny	1.7E+00	& \tiny	7.4E-01	& \tiny	1.8E-01	& \tiny	2.9E-03	& \tiny	7.2E-02	\\

\tiny	n4b1.50v30t405	& \tiny	7.5E-01	& \tiny	5.9E-01	& \tiny	2.9E-01	& \tiny	5.7E-02	& \tiny	1.1E-03	& \tiny	2.1E-02	\\
\tiny	n4b1.50v30t435	& \tiny	9.1E-01	& \tiny	6.8E-01	& \tiny	2.9E-01	& \tiny	7.2E-02	& \tiny	1.1E-03	& \tiny	2.8E-02	\\
\tiny	n4b1.50v30t825	& \tiny	6.4E+00	& \tiny	5.1E+00	& \tiny	2.2E+00	& \tiny	5.8E-01	& \tiny	1.0E-02	& \tiny	2.4E-01	\\
\\
\hline
\\
\tiny	n4b1.75v22t1115	& \tiny	3.0E+00	& \tiny	3.9E+00	& \tiny	3.4E+00	& \tiny	6.9E-01	& \tiny	4.5E-02	& \tiny	2.2E-01	\\
\tiny	n4b1.75v22t1180	& \tiny	3.5E+00	& \tiny	4.4E+00	& \tiny	3.8E+00	& \tiny	7.7E-01	& \tiny	4.8E-02	& \tiny	2.4E-01	\\
\tiny	n4b1.75v22t1260	& \tiny	4.1E+00	& \tiny	5.0E+00	& \tiny	4.1E+00	& \tiny	8.7E-01	& \tiny	5.6E-02	& \tiny	2.8E-01	\\

\tiny	n4b1.75v25t855	& \tiny	5.2E+00	& \tiny	6.5E+00	& \tiny	6.5E+00	& \tiny	1.6E+00	& \tiny	2.3E-01	& \tiny	7.8E-01	\\
\tiny	n4b1.75v25t885	& \tiny	5.4E+00	& \tiny	6.7E+00	& \tiny	6.7E+00	& \tiny	1.7E+00	& \tiny	2.4E-01	& \tiny	8.1E-01	\\
\tiny	n4b1.75v25t920	& \tiny	5.8E+00	& \tiny	7.1E+00	& \tiny	6.9E+00	& \tiny	1.8E+00	& \tiny	2.4E-01	& \tiny	8.3E-01	\\

\tiny	n4b1.75v28t660	& \tiny	9.7E-01	& \tiny	9.1E-01	& \tiny	5.9E-01	& \tiny	7.7E-02	& \tiny	2.1E-03	& \tiny	2.5E-02	\\
\tiny	n4b1.75v28t705	& \tiny	1.3E+00	& \tiny	1.1E+00	& \tiny	5.9E-01	& \tiny	1.0E-01	& \tiny	2.1E-03	& \tiny	3.7E-02	\\
\tiny	n4b1.75v28t890	& \tiny	9.6E+00	& \tiny	1.1E+01	& \tiny	1.0E+01	& \tiny	2.8E+00	& \tiny	4.2E-01	& \tiny	1.4E+00	\\

\tiny	n4b1.75v30t570	& \tiny	8.1E-01	& \tiny	6.3E-01	& \tiny	3.1E-01	& \tiny	6.0E-02	& \tiny	1.0E-03	& \tiny	2.2E-02	\\
\tiny	n4b1.75v30t600	& \tiny	1.1E+00	& \tiny	8.4E-01	& \tiny	3.7E-01	& \tiny	8.7E-02	& \tiny	1.3E-03	& \tiny	3.3E-02	\\
\tiny	n4b1.75v30t670	& \tiny	2.0E+00	& \tiny	1.5E+00	& \tiny	6.4E-01	& \tiny	1.7E-01	& \tiny	2.6E-03	& \tiny	6.7E-02	\\
\\
\hline
\\
\tiny	n4b2.00v28t835	& \tiny	1.1E+00	& \tiny	1.2E+00	& \tiny	9.3E-01	& \tiny	1.1E-01	& \tiny	3.6E-03	& \tiny	3.2E-02	\\
\tiny	n4b2.00v28t880	& \tiny	1.5E+00	& \tiny	1.6E+00	& \tiny	1.2E+00	& \tiny	1.6E-01	& \tiny	4.6E-03	& \tiny	4.8E-02	\\
\tiny	n4b2.00v28t920	& \tiny	3.1E+00	& \tiny	3.7E+00	& \tiny	2.8E+00	& \tiny	4.9E-01	& \tiny	1.7E-02	& \tiny	1.4E-01	\\

\tiny	n4b2.00v30t730	& \tiny	7.2E-01	& \tiny	5.8E-01	& \tiny	3.2E-01	& \tiny	5.0E-02	& \tiny	9.7E-04	& \tiny	1.7E-02	\\
\tiny	n4b2.00v30t755	& \tiny	9.5E-01	& \tiny	7.7E-01	& \tiny	4.0E-01	& \tiny	7.1E-02	& \tiny	1.3E-03	& \tiny	2.6E-02	\\
\tiny	n4b2.00v30t775	& \tiny	1.1E+00	& \tiny	8.6E-01	& \tiny	4.0E-01	& \tiny	8.6E-02	& \tiny	1.4E-03	& \tiny	3.2E-02	\\
\\
\hline
\\
\tiny	n5b0.30v10t16	& \tiny	1.1E+01	& \tiny	1.5E+01	& \tiny	1.1E+01	& \tiny	2.8E+00	& \tiny	1.9E-01	& \tiny	9.9E-01	\\
\tiny	n5b0.30v10t33	& \tiny	1.2E+01	& \tiny	1.7E+01	& \tiny	1.3E+01	& \tiny	3.4E+00	& \tiny	2.4E-01	& \tiny	1.2E+00	\\
\tiny	n5b0.30v10t67	& \tiny	1.3E+01	& \tiny	1.8E+01	& \tiny	1.4E+01	& \tiny	3.7E+00	& \tiny	2.7E-01	& \tiny	1.3E+00	\\

\tiny	n5b0.30v12t23	& \tiny	1.3E+01	& \tiny	1.9E+01	& \tiny	1.4E+01	& \tiny	4.7E+00	& \tiny	4.6E-01	& \tiny	1.8E+00	\\
\tiny	n5b0.30v12t48	& \tiny	1.6E+01	& \tiny	2.3E+01	& \tiny	1.7E+01	& \tiny	5.8E+00	& \tiny	5.8E-01	& \tiny	2.3E+00	\\
\tiny	n5b0.30v12t69	& \tiny	1.7E+01	& \tiny	2.3E+01	& \tiny	1.7E+01	& \tiny	5.6E+00	& \tiny	5.3E-01	& \tiny	2.2E+00	\\
\\
\hline
\\
\tiny	n5b0.45v10t51	& \tiny	1.1E+01	& \tiny	1.6E+01	& \tiny	1.2E+01	& \tiny	3.0E+00	& \tiny	2.0E-01	& \tiny	9.9E-01	\\
\tiny	n5b0.45v10t88	& \tiny	1.2E+01	& \tiny	1.6E+01	& \tiny	1.2E+01	& \tiny	3.2E+00	& \tiny	2.2E-01	& \tiny	1.0E+00	\\
\tiny	n5b0.45v10t115	& \tiny	1.2E+01	& \tiny	1.6E+01	& \tiny	1.3E+01	& \tiny	3.2E+00	& \tiny	2.2E-01	& \tiny	1.1E+00	\\

\tiny	n5b0.45v12t58	& \tiny	1.2E+01	& \tiny	1.7E+01	& \tiny	1.3E+01	& \tiny	4.1E+00	& \tiny	3.8E-01	& \tiny	1.5E+00	\\
\tiny	n5b0.45v12t76	& \tiny	1.3E+01	& \tiny	1.9E+01	& \tiny	1.4E+01	& \tiny	4.5E+00	& \tiny	4.2E-01	& \tiny	1.7E+00	\\

\tiny	n5b0.45v15t24	& \tiny	1.1E+01	& \tiny	1.6E+01	& \tiny	1.2E+01	& \tiny	4.5E+00	& \tiny	6.0E-01	& \tiny	2.0E+00	\\
\tiny	n5b0.45v15t40	& \tiny	2.0E+01	& \tiny	2.8E+01	& \tiny	2.2E+01	& \tiny	9.0E+00	& \tiny	1.6E+00	& \tiny	4.7E+00	\\
\tiny	n5b0.45v15t55	& \tiny	2.4E+01	& \tiny	3.3E+01	& \tiny	2.5E+01	& \tiny	1.0E+01	& \tiny	1.8E+00	& \tiny	5.4E+00	\\

\tiny	n5b0.45v18t13	& \tiny	1.3E+01	& \tiny	2.0E+01	& \tiny	1.5E+01	& \tiny	6.3E+00	& \tiny	1.0E+00	& \tiny	3.2E+00	\\
\tiny	n5b0.45v18t24	& \tiny	2.8E+01	& \tiny	4.3E+01	& \tiny	3.4E+01	& \tiny	1.7E+01	& \tiny	4.1E+00	& \tiny	1.1E+01	\\
\tiny	n5b0.45v18t35	& \tiny	3.8E+01	& \tiny	5.4E+01	& \tiny	4.3E+01	& \tiny	2.1E+01	& \tiny	5.3E+00	& \tiny	1.3E+01	\\
\\
\hline
\end{tabular}
\end{center}
\end{table*}

\setcounter{table}{1}
\begin{table*}
\begin{center}
\caption{\textit{(continued)} Predicted integrated intensities of the p-H$_2$O lines to be targeted by the HIFI and PACS receivers of the \textit{Herschel} telescope. The units are K km s$^{-1}$. The models considered are those which best fit the H$_2$ pure rotational excitation diagram at the position of  the SiO knot. A model referred to as \lq n4b0.45v15t335' has the following set of input parameters:~a pre-shock density of 10$^4$~cm$^{-3}$, a magnetic-field parameter of 0.45, a shock velocity of 15 km s$^{-1}$, and an age of 335 years.}            
\label{tableb2}      
\begin{tabular}{c c c c c c c c c c c c c }   
\hline                  
\hline
\tiny	transition name	& \tiny	1$_{\rm{11}}$ - 0$_{\rm{00}}$	& \tiny	2$_{\rm{02}}$ - 1$_{\rm{11}}$	& \tiny	2$_{\rm{11}}$ - 2$_{\rm{02}}$	& \tiny	3$_{\rm{13}}$ - 2$_{\rm{02}}$	& \tiny	4$_{\rm{04}}$ - 3$_{\rm{13}}$	& \tiny	3$_{\rm{22}}$ - 2$_{\rm{11}}$	\\
\tiny	frequency (GHz)	& \tiny	1113.343	& \tiny	987.927	& \tiny	752.033	& \tiny	2164.132	& \tiny	2391.573	& \tiny	3331.458	\\
\hline														
\hline
\\	
\tiny	n5b0.60v10t52	& \tiny	1.0E+01	& \tiny	1.4E+01	& \tiny	1.1E+01	& \tiny	2.5E+00	& \tiny	1.5E-01	& \tiny	8.1E-01	\\
\tiny	n5b0.60v10t105	& \tiny	1.1E+01	& \tiny	1.5E+01	& \tiny	1.1E+01	& \tiny	2.8E+00	& \tiny	1.8E-01	& \tiny	8.9E-01	\\
\tiny	n5b0.60v10t155	& \tiny	1.1E+01	& \tiny	1.5E+01	& \tiny	1.2E+01	& \tiny	2.9E+00	& \tiny	1.8E-01	& \tiny	9.1E-01	\\

\tiny	n5b0.60v12t77	& \tiny	1.0E+01	& \tiny	1.4E+01	& \tiny	1.1E+01	& \tiny	3.1E+00	& \tiny	2.4E-01	& \tiny	1.1E+00	\\
\tiny	n5b0.60v12t85	& \tiny	1.0E+01	& \tiny	1.4E+01	& \tiny	1.1E+01	& \tiny	3.2E+00	& \tiny	2.5E-01	& \tiny	1.1E+00	\\
\tiny	n5b0.60v12t120	& \tiny	1.2E+01	& \tiny	1.7E+01	& \tiny	1.3E+01	& \tiny	3.9E+00	& \tiny	3.4E-01	& \tiny	1.4E+00	\\

\tiny	n5b0.60v15t53	& \tiny	1.1E+01	& \tiny	1.6E+01	& \tiny	1.3E+01	& \tiny	4.7E+00	& \tiny	6.8E-01	& \tiny	2.1E+00	\\
\tiny	n5b0.60v15t72	& \tiny	1.9E+01	& \tiny	2.6E+01	& \tiny	2.1E+01	& \tiny	8.1E+00	& \tiny	1.5E+00	& \tiny	4.2E+00	\\

\tiny	n5b0.60v20t26	& \tiny	8.4E+00	& \tiny	1.4E+01	& \tiny	1.1E+01	& \tiny	5.1E+00	& \tiny	1.1E+00	& \tiny	3.0E+00	\\
\tiny	n5b0.60v20t34	& \tiny	2.4E+01	& \tiny	3.4E+01	& \tiny	2.8E+01	& \tiny	1.4E+01	& \tiny	4.4E+00	& \tiny	1.0E+01	\\
\tiny	n5b0.60v20t62	& \tiny	5.1E+01	& \tiny	6.9E+01	& \tiny	5.5E+01	& \tiny	2.7E+01	& \tiny	8.3E+00	& \tiny	1.9E+01	\\
\\
\hline
\\
\tiny	n5b0.75v10t115	& \tiny	1.0E+01	& \tiny	1.4E+01	& \tiny	1.1E+01	& \tiny	2.5E+00	& \tiny	1.5E-01	& \tiny	7.7E-01	\\
\tiny	n5b0.75v10t220	& \tiny	1.0E+01	& \tiny	1.4E+01	& \tiny	1.1E+01	& \tiny	2.6E+00	& \tiny	1.6E-01	& \tiny	8.1E-01	\\
\tiny	n5b0.75v10t250	& \tiny	1.0E+01	& \tiny	1.4E+01	& \tiny	1.1E+01	& \tiny	2.6E+00	& \tiny	1.6E-01	& \tiny	8.1E-01	\\

\tiny	n5b0.75v12t100	& \tiny	9.7E+00	& \tiny	1.4E+01	& \tiny	1.1E+01	& \tiny	2.9E+00	& \tiny	2.1E-01	& \tiny	9.7E-01	\\
\tiny	n5b0.75v12t140	& \tiny	9.8E+00	& \tiny	1.4E+01	& \tiny	1.1E+01	& \tiny	3.0E+00	& \tiny	2.2E-01	& \tiny	1.0E+00	\\
\tiny	n5b0.75v12t165	& \tiny	1.0E+01	& \tiny	1.5E+01	& \tiny	1.1E+01	& \tiny	3.2E+00	& \tiny	2.5E-01	& \tiny	1.1E+00	\\

\tiny	n5b0.75v15t86	& \tiny	1.0E+01	& \tiny	1.5E+01	& \tiny	1.2E+01	& \tiny	4.1E+00	& \tiny	5.6E-01	& \tiny	1.8E+00	\\
\tiny	n5b0.75v15t104	& \tiny	1.6E+01	& \tiny	2.2E+01	& \tiny	1.8E+01	& \tiny	6.7E+00	& \tiny	1.1E+00	& \tiny	3.3E+00	\\
\\
\hline
\\
\tiny	n5b1.00v10t385	& \tiny	9.4E+00	& \tiny	1.3E+01	& \tiny	1.0E+01	& \tiny	2.3E+00	& \tiny	1.3E-01	& \tiny	6.8E-01	\\
\tiny	n5b1.00v10t410	& \tiny	9.4E+00	& \tiny	1.3E+01	& \tiny	1.0E+01	& \tiny	2.3E+00	& \tiny	1.3E-01	& \tiny	6.8E-01	\\

\tiny	n5b1.00v12t205	& \tiny	9.6E+00	& \tiny	1.4E+01	& \tiny	1.1E+01	& \tiny	2.7E+00	& \tiny	1.9E-01	& \tiny	8.9E-01	\\
\tiny	n5b1.00v12t235	& \tiny	9.5E+00	& \tiny	1.3E+01	& \tiny	1.0E+01	& \tiny	2.7E+00	& \tiny	1.9E-01	& \tiny	8.9E-01	\\
\tiny	n5b1.00v12t255	& \tiny	9.6E+00	& \tiny	1.4E+01	& \tiny	1.0E+01	& \tiny	2.7E+00	& \tiny	1.9E-01	& \tiny	9.0E-01	\\

\tiny	n5b1.00v15t160	& \tiny	9.6E+00	& \tiny	1.4E+01	& \tiny	1.1E+01	& \tiny	3.6E+00	& \tiny	4.4E-01	& \tiny	1.5E+00	\\
\tiny	n5b1.00v15t175	& \tiny	1.2E+01	& \tiny	1.6E+01	& \tiny	1.3E+01	& \tiny	4.6E+00	& \tiny	6.4E-01	& \tiny	2.0E+00	\\
\\
\hline
\\
\tiny	n5b1.25v15t205	& \tiny	8.7E+00	& \tiny	1.2E+01	& \tiny	9.8E+00	& \tiny	3.0E+00	& \tiny	3.0E-01	& \tiny	1.1E+00	\\
\tiny	n5b1.25v15t265	& \tiny	1.0E+01	& \tiny	1.4E+01	& \tiny	1.2E+01	& \tiny	3.8E+00	& \tiny	4.5E-01	& \tiny	1.5E+00	\\
\\
\hline
\\
\tiny	n5b1.50v10t525	& \tiny	8.4E+00	& \tiny	1.1E+01	& \tiny	9.2E+00	& \tiny	1.8E+00	& \tiny	8.6E-02	& \tiny	5.0E-01	\\
\tiny	n5b1.50v10t775	& \tiny	8.4E+00	& \tiny	1.1E+01	& \tiny	9.2E+00	& \tiny	1.8E+00	& \tiny	8.8E-02	& \tiny	5.0E-01	\\

\tiny	n5b1.50v15t335	& \tiny	8.4E+00	& \tiny	1.2E+01	& \tiny	9.5E+00	& \tiny	2.8E+00	& \tiny	2.6E-01	& \tiny	1.0E+00	\\
\tiny	n5b1.50v15t370	& \tiny	8.7E+00	& \tiny	1.2E+01	& \tiny	9.9E+00	& \tiny	3.0E+00	& \tiny	2.8E-01	& \tiny	1.1E+00	\\
\\
\hline
\\
\tiny	n5b1.75v10t505	& \tiny	7.9E+00	& \tiny	1.1E+01	& \tiny	8.7E+00	& \tiny	1.6E+00	& \tiny	7.0E-02	& \tiny	4.3E-01	\\
\tiny	n5b1.75v10t915	& \tiny	7.9E+00	& \tiny	1.1E+01	& \tiny	8.8E+00	& \tiny	1.6E+00	& \tiny	7.3E-02	& \tiny	4.4E-01	\\

\tiny	n5b1.75v12t315	& \tiny	8.3E+00	& \tiny	1.2E+01	& \tiny	9.3E+00	& \tiny	2.6E+00	& \tiny	2.1E-01	& \tiny	8.9E-01	\\
\tiny	n5b1.75v12t410	& \tiny	8.3E+00	& \tiny	1.2E+01	& \tiny	9.4E+00	& \tiny	2.6E+00	& \tiny	2.2E-01	& \tiny	9.1E-01	\\
\tiny	n5b1.75v12t465	& \tiny	8.3E+00	& \tiny	1.2E+01	& \tiny	9.4E+00	& \tiny	2.6E+00	& \tiny	2.2E-01	& \tiny	9.2E-01	\\

\tiny	n5b1.75v18t305	& \tiny	9.4E+00	& \tiny	1.3E+01	& \tiny	1.1E+01	& \tiny	3.7E+00	& \tiny	5.7E-01	& \tiny	1.7E+00	\\
\tiny	n5b1.75v18t325	& \tiny	9.5E+00	& \tiny	1.3E+01	& \tiny	1.1E+01	& \tiny	3.7E+00	& \tiny	5.8E-01	& \tiny	1.8E+00	\\
\tiny	n5b1.75v18t340	& \tiny	1.0E+01	& \tiny	1.4E+01	& \tiny	1.2E+01	& \tiny	4.1E+00	& \tiny	6.8E-01	& \tiny	2.0E+00	\\
\\
\hline
\\
\tiny	n5b2.00v18t435	& \tiny	8.8E+00	& \tiny	1.2E+01	& \tiny	1.0E+01	& \tiny	3.3E+00	& \tiny	4.5E-01	& \tiny	1.4E+00	\\
\tiny	n5b2.00v18t440	& \tiny	9.0E+00	& \tiny	1.2E+01	& \tiny	1.1E+01	& \tiny	3.4E+00	& \tiny	4.7E-01	& \tiny	1.5E+00	
\\
\hline
\end{tabular}
\end{center}
\end{table*}

\end{appendix}

\end{document}